\documentclass[11pt,a4paper]{article}
\pdfoutput=1

\usepackage{jcappub}

\usepackage{inputenc}
\usepackage{subfigure}
\usepackage{multirow}

\inputencoding{latin9}

\newcommand{\nc}{\newcommand}

\nc{\ie}{{\em i.e. }}
\nc{\eg}{{\em e.g. }}

\begin{document}

\title{PyCOOL - a Cosmological Object-Oriented Lattice code written in Python}

\author[a,b]{J. Sainio}
\emailAdd{jani.sainio@utu.fi}
\emailAdd{jtksai@gmail.com}
\affiliation[a]{Turku School of Economics, University of Turku, FIN-20014 Turku, FINLAND}
\affiliation[b]{Department of Physics and Astronomy, University of Turku, FIN-20014 Turku, FINLAND}


\abstract{

There are a number of different phenomena in the early universe that have to be studied numerically with lattice simulations.
This paper presents a graphics processing unit (GPU) accelerated Python program called PyCOOL that solves the evolution of scalar fields in a lattice with very precise symplectic integrators.
The program has been written with the intention to hit a sweet spot of speed, accuracy and user friendliness. This has been achieved by using the Python language with the PyCUDA interface to make a program that is easy to adapt to different scalar field models.
In this paper we derive the symplectic dynamics that govern the evolution of the system and then present the implementation of the program in Python and PyCUDA. The functionality of the program is tested in a chaotic inflation preheating model, a single field oscillon case and in a supersymmetric curvaton model which leads to Q-ball production.
We have also compared the performance of a consumer graphics card to a professional Tesla compute card in these simulations.
We find that the program is not only accurate but also very fast.
To further increase the usefulness of the program we have equipped it with numerous post-processing functions that provide useful information about the cosmological model. These include various spectra and statistics of the fields. The program can be additionally used to calculate the generated curvature perturbation.
The program is publicly available under GNU General Public License at \href{https://github.com/jtksai/PyCOOL}{https://github.com/jtksai/PyCOOL}. Some additional information can be found from \href{http://www.physics.utu.fi/tiedostot/theory/particlecosmology/pycool/}{http://www.physics.utu.fi/tiedostot/theory/particlecosmology/pycool/}.}

\maketitle

\section{Introduction}

Recent developments in computer technology have made graphic processing units (GPUs) an increasingly popular way to solve computationally difficult problems \cite{Belleman:2007kv,Brunner:2007sy,Zwart:2008bp,Ishikawa:2008pf,Szalay:2008,Anselmi:2008hi,Ford:2008em,Ord:2009xk,Gaburov:2009dy,
Demchik:2009ni,Chung:2009yb,Schive:2009hw,Jonsson:2009dh,Groen:2009,Nakasato:2009xq,Khanna:2009zs,Hagiwara:2009cy,
CapuzzoDolcetta:2009er,Banerjee:2009hs,Wang:2009st,Januszewski:2009}. 
The shear floating point power of GPUs is however useless if it cannot be harnessed effectively. The use of a suitable language is therefore of paramount importance.
During the last years the programming of modern GPUs has thankfully evolved drastically and it has become relatively easy. NVIDIA's CUDA architecture \cite{NVIDIA:2011} and the OpenCL language \cite{OpenCL:2011} of the Khronos Group are very similar to C syntax and therefore lower considerably the learning curve needed to write programs that use GPUs as co-processors. In addition to these there are now multiple other publicly available computer languages that can be used to call GPUs directly or through one of the two aforementioned languages. PyCUDA \cite{Klockner:2009} is an interesting and a powerful interface to NVIDIA's CUDA that is based on the Python scripting language. Together with its textual templating functionality it allows to easily modify the CUDA kernels which make it possible to make highly dynamical GPGPU programs that can be changed even at runtime.

Cosmology offers many computationally demanding problems \cite{Hoerner:1960,Hoerner:1963,Bertschinger:1998} that can be used to test and to develop current theories of particle physics.
The preheating phase of the post-inflationary universe is an interesting subject that has been studied in a myriad of papers (see for example \cite{Felder:2008}).
The non-linear dynamics that govern the evolution of scalar fields during this process have to be solved numerically. Fortunately there are a number of publicly available programs that designed for this. These include  LATTICEEASY \cite{Felder:2000hq}, DEFROST \cite{Frolov:2008hy}, PSpectRe \cite{Easther:2010qz} and HLATTICE \cite{Huang:2011gf} which are all C, C++ or Fortran based programs that use CPU(s) to evolve the scalar fields.

The lattice calculations are relatively easily parallelizable and therefore suitable for also GPU computing.
The only program to accomplish this up to this point has been CUDAEASY \cite{Sainio:2009hm} that was in essence an upgraded version of LATTICEEASY retrofitted to the specifications and demands of the GPU. Although it was shown to be much faster than the rest of the preheating codes it had many shortcomings that made the use of it unappealing. These include the fact that in order to change to a different scalar field model the CUDA kernels would have to be edited significantly. The program was also limited to single precision accuracy which can be a limiting factor in the simulations of the early universe.

Instead of only simply improving the existing code we decided to create a completely new program from scratch that takes the best parts of CUDAEASY and eradicates all of the known shortcomings. We call this program PyCOOL. The user friendliness of the software was made a priority in the development process in order to make a system that allows the user to concentrate on the physics instead of rewriting code. This has been accomplished by writing an extensive object-oriented Python program that uses PyCUDA to create the necessary CUDA kernels needed in the evolution steps and in the post-processing phase.


The integration system has also been changed completely. Whereas CUDAEASY used a second-order leap-frog method to evolve the fields PyCOOL is based on an entirely new symplectic method developed by A. Frolov and Z. Huang \cite{Frolov} which is also somewhat related to the integrator used in HLATTICE \cite{Huang:2011gf}. This symplectic algorithm is found to be especially suitable for the GPU since it does not need any auxiliary fields and hence has much smaller memory footprint compared for example to an ordinary fourth order Runge-Kutta algorithm.

One of the weaknesses of CUDAEASY was also the fact that it could only solve the evolution of the system.
However when analyzing the simulation results variables derived from the simulation data provide often more valuable information about the system.
We have therefore included numerous post processing functions in the program that can calculate various spectra used in LATTICEEASY and Defrost programs, statistics and effective masses of the fields. All of the results are currently stored in SILO files which can be easily visualized in the VisIt software. Many of the variables are also written in csv (comma separated values) format in case other visualization programs are preferred. On top of these we have also incorporated the ability to calculate curvature perturbations generated in during the evolution of the system.

The goal of this paper is to derive the symplectic dynamics that govern the evolution of the system and then to present the implementation of the program in Python and PyCUDA. It is organized as follows. In section II we present and derive the symplectic equations of motion in the linear and non-linear cases. We also present the symplectic integration methods used and the implementation of these integrators in PyCUDA. In section III we show the numerical results in chaotic inflation, oscillon and curvaton models. We conclude with a discussion in section IV.

\section{Symplectic Dynamics}

\subsection{Discretization of the System}

We will start by presenting the symplectic method of solving evolution of the system. Symplectic integrators are often used when the conservation of a physical quantity is paramount to the validity of the solutions: for example in the case of celestial mechanics long time integrations of the solar system are often done with different symplectic integrators.

The symplectic method of solving the dynamics during preheating was first developed by A. Frolov and Z. Huang \cite{Frolov} and we will follow their method closely. Instead of discretizing the equations of motion the starting point is to first discretize the action of the system spatially and then to derive the Hamiltonian equations of motion from this discrete action by variation. This method follows closely the principles used in different variational integrators.


The action from which the equations of motion are derived in general relativity is
\begin{equation} \label{eq:act}
S = \int \Big(-\frac{\mathcal{R}}{16 \pi G} + \mathcal{L}_m\Big)\sqrt{-g} \, \mathrm{d}^4 x,
\end{equation}
where $\mathcal{R}$ is the Ricci curvature scalar and $\mathcal{L}_m$ is the Lagrange function of the matter content of the universe.
The preheating phase of the universe is often modeled with semi-classical scalar fields that are minimally coupled to gravity and interact with each other through a potential term.
By demanding Lorentz invariance $\mathcal{L}_m$ is compelled to be a function of only the fields and their temporal and spatial first derivatives and action (\ref{eq:act}) then reads
\begin{equation} \label{eq:act2}
S = \int \Big(-\frac{m_{Pl}^2}{2} \mathcal{R} + \sum_i \frac{1}{2}\partial_{\mu}\phi_{i} \partial^{\mu}\phi_{i} - V(\phi_{1},...,\phi_{N})\Big)\sqrt{-g} \, \mathrm{d}^4 x,
\end{equation}
where index $i=1,...,N$ labels the different fields and we have defined the reduced Planck mass $m_{Pl} = 1/\sqrt{8 \pi G}$.

We will assume that the space-time is spatially homogeneous, isotropic and flat \ie a Friedmann-Robertson-Walker universe. The space-time interval can be then written as
\begin{equation} \label{eq:metric}
ds^2 = a(\eta)^2(d\eta^2 - d \vec{x} ^2),
\end{equation}
where $a(\eta)$ is the scale parameter and we have used conformal time $\eta$ which makes the Hamiltonian equations of motion simpler compared to physical time. From this metric it follows that the determinant of the metric tensor equals
\begin{equation}
\sqrt{-g} = a^4
\end{equation}
and the Ricci scalar reads
\begin{equation}
\mathcal{R} = -6 \frac{a''}{a^3},
\end{equation}
where prime denotes differentiation with respect to conformal time.


The action now reads
\begin{equation} \label{eq:act3}
S = \int \Bigg(-3 a'^2 m_{Pl}^2 V_L (dx)^3 + \int \Big( \sum_i \frac{\phi_{i}'^2}{2 a^2} + \frac{\phi_{i}\nabla^2\phi_{i}}{2 a^2} - V(\phi_{1},...,\phi_{N})\Big) a^4 \mathrm{d}^3 x\Bigg)\mathrm{d}\eta ,
\end{equation}
where we have done partial integration with respect to time for the scale parameter term and also the integrated over the spatial coordinates. Note that $V_L (\mathrm{d}x)^3$ equals the volume of the periodic lattice in which the fields are embedded and $\nabla$ is the gradient operator with respect to comoving coordinates. 
In order to optimize the GPU implementation we have also done a spatial partial integration to the gradient terms of the scalar field which was also used in CUDAEASY \cite{Sainio:2009hm}.
This way the computationally difficult $(\nabla \phi)^2 $ terms can be eliminated from the evolution equations.  

Next step is to make the spatial discretization of the lattice where the system in transformed into $i \times n^3$ time dependent scalar fields that are coupled to each other. In the discretization we have used the second order accurate stencils for the gradient terms presented in \cite{Patra:2005} which were also used in DEFROST \cite{Frolov:2008hy}, CUDAEASY \cite{Sainio:2009hm} and in the symplectic method of Frolov and Huang \cite{Frolov}. In this notation the Laplacian operator reads
\begin{equation} \label{eq:D}
D[\phi](x,y,z) = \sum^{x+1}_{i=x-1}\sum^{y+1}_{j=y-1}\sum^{z+1}_{k=z-1} c_{d(i,j,k)}\phi(i,j,k) = \sum_{\alpha} c_{d(\alpha)}\phi(\alpha),
\end{equation}
where $c_{d(\alpha)}$ are the discretization coefficients of the Laplacian \cite{Patra:2005}.
We also have
\begin{equation} \label{eq:G}
G[\phi](x,y,z) = \frac{1}{2} \sum_{\alpha} c_{d(\alpha)}\Big(\phi(\alpha) - \phi(x,y,z)\Big)^2
\end{equation}
for $(\nabla \phi)^2$ terms used in the energy density calculations. Huang has also implemented a fourth order accurate stencil in HLATTICE program \cite{Huang:2011gf}. Although the implementation of this in CUDA is possible it would however demand rewriting of the current CUDA code and this is left for future upgrades of PyCOOL.

With the spatial discretization integrals transform to summations over the grid
\begin{equation}
\int \mathrm{d}^3 x \rightarrow \sum_{x,y,z} dx^3
\end{equation}
and the action (\ref{eq:act3}) reads
\begin{equation} \label{eq:act-disc}
\begin{aligned}
S = & (\mathrm{d}x)^3 \int \Bigg[-3 a'^2 V_L m_{Pl}^2 + \sum_{\vec{x}} a^2 \Big( \sum_i \Big(\frac{\phi_{i,\vec{x}}'^2}{2}
+ \frac{\phi_{i,\vec{x}}D[\phi_{i,\vec{x}}](\vec{x})}{2 dx^2}\Big) - a^2 V(\phi_{1,\vec{x}},...,\phi_{N,\vec{x}})\Big)\Bigg]\mathrm{d}\eta \\
  = & (\mathrm{d}x)^3 \int \mathcal{L}[a,\phi_{1},...\phi_{N} ]\mathrm{d}\eta ,
\end{aligned}
\end{equation}
where we have labeled the different scalar fields with the index $i$ and their location in the lattice $\vec{x}$.

In order to determine the Hamiltonian $\mathcal{H}$ the canonical momentum variables have to be determined. From the Lagrangian $\mathcal{L}$ we get for the scale parameter
\begin{equation} \label{eq:p-a}
p_a = \frac{\partial \mathcal{L}}{\partial a'} = -6 a' V_L m_{Pl}^2
\end{equation}
and for the field variables
\begin{equation} \label{eq:pi-i}
\pi_{i,\vec{x}} = \frac{\partial \mathcal{L}}{\partial \phi'_{i,\vec{x}}} = a^2 \phi'_{i,\vec{x}}.
\end{equation}
After a simple Legendre transformation of the Lagrangian function the Hamiltonian function of the system equals
\begin{equation} \label{eq:H}
\mathcal{H} = -\frac{ p_a^2 }{12 V_L m_{Pl}^2} + \sum_{i,\vec{x}} a^4 \Bigg( \frac{\pi_{i,\vec{x}}^2}{2 a^6} - \frac{\phi_{i,\vec{x}}D[\phi_{i,\vec{x}}](\vec{x})}{ 2 a^2 dx^2} + V(\phi_{1,\vec{x}},...,\phi_{N,\vec{x}})\Bigg).
\end{equation}

The Hamiltonian equations related to this Hamiltonian now read for the scale parameter
\begin{equation} \label{eq:a}
\begin{aligned}
a' &= \frac{\partial \mathcal{H}}{\partial p_a} = - \frac{p_a}{ 6 V_L m_{Pl}^2}\\
p_a' &= - \frac{\partial \mathcal{H}}{\partial a} = \sum_{i,\vec{x}} a^3 \Bigg( \frac{\pi_{i,\vec{x}}^2}{a^6} + \frac{\phi_{i,\vec{x}}D[\phi_{i,\vec{x}}](\vec{x})}{ a^2 dx^2} - 4 V(\phi_{1,\vec{x}},...,\phi_{N,\vec{x}})\Bigg).
\end{aligned}
\end{equation}
Similarly the equations of motion of scalar field $i$ at grid point $\vec{z}$ read
\begin{equation} \label{eq:phi}
\begin{aligned}
\phi_{i,\vec{z}}' &= \frac{\partial \mathcal{H}}{\partial (\pi_{i,\vec{z}})} = \frac{\pi_{i,\vec{z}}}{a^2} \\
\pi_{i,\vec{z}}' &= -\frac{\partial \mathcal{H}}{\partial (\phi_{i,\vec{z}})} =  a^2 \frac{D[\phi_{i,\vec{z}}](\vec{z})}{ dx^2} - a^4 \frac{\partial V }{\partial (\phi_{i,\vec{z}})}
\end{aligned}
\end{equation}
which follow from equation (\ref{eq:H}) by differentiating under the summation sign and by summing over the coefficients $c_{d(\alpha)}$.

By noting that the total energy density and pressure of the scalar fields are given by the usual formulas
\begin{equation} \label{eq:rho-pres}
\begin{aligned}
\rho &= \sum_{i}\Big(\frac{(\phi'_{i})^2}{2 a^2} + \frac{1}{2a^2}(\nabla\phi_{i})^2\Big) + V(\phi_{i})\\
P &= \sum_{i}\Big(\frac{(\phi'_{i})^2}{2 a^2} - \frac{1}{6a^2}(\nabla\phi_{i})^2\Big) - V(\phi_{i})\\
\end{aligned}
\end{equation}
and by remembering that in the periodic lattice as a result of the Stoke's theorem
\begin{equation}
\sum_{\vec{x}} \frac{\phi_{i,\vec{x}}D[\phi_{i,\vec{x}}](\vec{x})}{ dx^2} = - \sum_{\vec{x}} \frac{G[\phi_{i,\vec{x}}](\vec{x})}{ dx^2}
\end{equation}
it can be easily seen that the right hand side of the Hamiltonian (\ref{eq:H}) is actually a discretized version of the first Friedmann equation
\begin{equation} \label{eq:Fried-1}
(a')^2 = \frac{a^4}{3 m_{Pl}^2}\langle\rho\rangle,
\end{equation}
where $\langle\rangle$ denotes average over the volume of the lattice. Similarly it can be easily seen that the equation of motion of $p_{a}$ is a discretized version of the second Friedmann equation
\begin{equation} \label{eq:Fried-2}
a'' = -\frac{a^3}{6 m_{Pl}^2}\langle\rho-3P\rangle.
\end{equation}

In the curvaton scenario and during curvaton decay a homogeneous radiation component originating from the inflaton fields is  included often in the system. We have therefore modified the Hamiltonian 
(\ref{eq:H}) into form
\begin{equation} \label{eq:H2}
\mathcal{H} = -\frac{ p_a^2 }{12 V_L m_{Pl}^2} + \sum_{i,\vec{x}} a^4 \left( \frac{\pi_{i,\vec{x}}^2}{2 a^6} - \frac{\phi_{i,\vec{x}}D[\phi_{i,\vec{x}}](\vec{x})}{ 2 a^2 dx^2} + V(\phi_{1,\vec{x}},...,\phi_{N,\vec{x}})\right) + a^4\left(\frac{ V_L \rho_{\gamma,0}}{a^4} + \frac{ V_L \rho_{m,0}}{a^3}\right),
\end{equation}
where we have now incorporated homogeneous radiation and non-relativistic matter components into the system. Since the scale parameters in the radiation component now cancel out the effect of radiation is only seen in the value of $\mathcal{H}$ but it does not show in the equations of motion of the system, which is consistent with the second Friedmann equation (\ref{eq:Fried-2}) for radiation.
The homogeneous matter component leads to an additional $V_L\rho_{m,0}$ term in the equation of motion of $p_a$ which is again consistent with the second Friedmann equation (\ref{eq:Fried-2}) for non-relativistic matter.

\subsection{Linearized equations}

Solving numerically the evolution generated by Hamiltonian (\ref{eq:H2}) is computationally very heavy due to the number of equations involved, $i \times n^3$. However as noted in \cite{Chambers:2009ki} the dynamics of the scalar fields (\ref{eq:phi}) can be approximated in some cases with linearized equations which make the numerical calculations much faster. We will therefore derive also the Hamiltonian equations related to this linear case by first separating the field into a homogeneous and a non-homogeneous perturbation component. By noticing that the Fourier modes of these linear perturbations do not couple to each other we can reduce the number of solved equations drastically and therefore solve the evolution faster.

We decompose the scalar fields into a homogeneous and non-homogeneous perturbation part
\begin{equation} \label{eq:phi-lin}
\phi_{i}(\vec{x}) = \phi_{\textrm{0},i} + \delta\phi_{i}(\vec{x})
\end{equation}
where the sum of the perturbation terms over the lattice is assumed to vanish \ie
\begin{equation}
\sum_{\vec{x}} \delta \phi_{i}(\vec{x}) = 0.
\end{equation}
In the code this constraint is satisfied by making sure that the homogeneous $k_\textrm{eff}^2 = 0$ mode of the perturbed variables stays close to zero at all times.
The Lagrangian (\ref{eq:act-disc})
\begin{equation}
\mathcal{L} = -3 a'^2 V_L m_{Pl}^2 + \sum_{\vec{x}} a^2 \Big( \sum_i  \Big(\frac{\phi_{i,\vec{x}}'^2}{2} - \frac{G[\phi_{i,\vec{x}}](\vec{x})}{ 2 dx^2}\Big) + a^2 V(\phi_{1,\vec{x}},...,\phi_{N,\vec{x}})\Big)
\end{equation}
factorizes into
\begin{equation} \label{eq:L-pert}
\begin{aligned}
\mathcal{L} = & -3 a'^2 V_L m_{Pl}^2 + a^2 V_L\Big( \sum_i \frac{1}{2} \Big(\phi_{\textrm{0},i}' \Big)^2 + a^2 V(\phi_{\textrm{0},1},...,\phi_{\textrm{0},N})\Big)\\
& + \sum_{\vec{x}} a^2  \Big( \sum_i \Big(\phi_{\textrm{0},i}'\delta\phi_{i,\vec{x}}'\Big) + a^2 V^{(1)}(\phi_{1,\vec{x}},...,\phi_{N,\vec{x}})\Big)\\
& + \sum_{\vec{x}} a^2  \Big( \sum_i \Big(\frac{\delta\phi_{i,\vec{x}}'^2}{2} - \frac{G[\delta\phi_{i,\vec{x}}](\vec{x})}{ 2 dx^2}\Big) + a^2 V^{(2)}(\phi_{1,\vec{x}},...,\phi_{N,\vec{x}})\Big),\\
\end{aligned}
\end{equation}
where
\begin{equation} \label{eq:V-1}
 V^{(1)}(\phi_{1,\vec{x}},...,\phi_{N,\vec{x}}) = \frac{dV(\phi_{\textrm{0},1} + \epsilon\delta\phi_{1}(\vec{x}),...,\phi_{\textrm{0},N} + \epsilon\delta\phi_{N}(\vec{x}))}{d\epsilon}\Big|_{\epsilon=0}
\end{equation}
and
\begin{equation} \label{eq:V-2}
 V^{(2)}(\phi_{1,\vec{x}},...,\phi_{N,\vec{x}}) = \frac{1}{2}\frac{d^2V(\phi_{\textrm{0},1} + \epsilon\delta\phi_{1}(\vec{x}),...,\phi_{\textrm{0},N} + \epsilon\delta\phi_{N}(\vec{x}))}{d\epsilon ^2}\Big|_{\epsilon=0}
\end{equation}
are the potential terms at first and second order in the perturbed variables, respectively.
For example for a simple polynomial potential function $V = \frac{1}{2}m_{\phi}^2 \phi^2 + \frac{1}{2} g^2 \phi^2 \chi^2$ the second order term reads
\begin{equation}
 V^{(2)}(\phi,\chi) = \frac{1}{2}m_{\phi}^2 \delta\phi^2 + \frac{1}{2} g^2 \delta\phi^2 \chi_0^2 + 2 g^2 \phi_0 \chi_0 \delta\phi \delta\chi + \frac{1}{2} g^2 \phi_0^2 \delta\chi^2.
\end{equation}

It is now easy to write the Lagrangians related to different orders of the field perturbations. The background Lagrangian simply reads
\begin{equation}
\mathcal{L}^{\textrm{(0)}} =  -3 a'^2 V_L m_{Pl}^2 + a^2 V_L\Big( \sum_i \frac{1}{2} \Big(\phi_{\textrm{0},i}' \Big)^2 + a^2 V(\phi_{\textrm{0},1},...,\phi_{\textrm{0},N})\Big).
\end{equation}
At first order the Lagrangian term can be simplified by partial integrating $a^2 \phi_{\textrm{0},i}'\delta\phi_{i,\vec{x}}'$ term in the corresponding action integral and
by writing the potential term (\ref{eq:V-1}) open after which the Lagrangian is
\begin{equation}
\mathcal{L}^{\textrm{(1)}} = \sum_{\vec{x},i} \Big( -2 a a'\phi_{\textrm{0},i}' - a^2\phi_{\textrm{0},i}'' + a^2 \frac{\partial V}{\partial (\phi_{\textrm{0},i})}\Big)\delta\phi_{i,\vec{x}}.
\end{equation}
By noticing that the terms inside the brackets equal the Euler-Lagrange equation of motion of the background field $i$ the first order Lagrangian terms equal zero.
At second order the Lagrangian can be easily read directly from (\ref{eq:L-pert})
\begin{equation}
\begin{aligned}
\mathcal{L}^{\textrm{(2)}} = \sum_{\vec{x}} a^2  \Big( \sum_i \Big(\frac{\delta\phi_{i,\vec{x}}'^2}{2} - \frac{G[\delta\phi_{i,\vec{x}}](\vec{x})}{ 2 dx^2}\Big) + a^2 V^{(2)}(\phi_{1,\vec{x}},...,\phi_{N,\vec{x}})\Big).\\
\end{aligned}
\end{equation}

In order to derive the Hamiltonian equations of motion the necessary canonical momentums have to be first defined. 
For the scale parameter we get
\begin{equation}
p_a = \frac{\partial \mathcal{L}}{\partial a'} = -6 a' V_L m_{Pl}^2
\end{equation}
and for the background field variables
\begin{equation} \label{eq:pi-bg}
\pi_{\textrm{0},i} = \frac{\partial \mathcal{L}}{\partial (\phi'_{\textrm{0},i})} = a^2 \phi'_{\textrm{0},i}.
\end{equation}
and finally for the field perturbations
\begin{equation} \label{eq:pi-1}
\delta\pi_{i,\vec{x}} = \frac{\partial \mathcal{L}}{\partial (\delta\phi_{i,\vec{x}}')} = a^2 \delta\phi_{i,\vec{x}}'.
\end{equation}

By doing again a Legendre transformation we get the Hamiltonian function of the perturbed system
\begin{equation} \label{eq:H-lin}
\begin{aligned}
\mathcal{H} = \mathcal{H}^{\textrm{(0)}} + \mathcal{H}^{\textrm{(2)}} =  & -\frac{ p_a^2 }{12 V_L m_{Pl}^2} + \sum_{i} a^4 V_L \Bigg( \frac{\pi_{\textrm{0},i}^2}{2 a^6} + V^ {(0)}(\phi_{1,\vec{x}},...,\phi_{N,\vec{x}})\Bigg)\\
& + \sum_{i,\vec{x}} a^4 \Bigg( \frac{\delta\pi_{i,\vec{x}}^2}{2 a^6} - \frac{\delta\phi_{i,\vec{x}}D[\delta\phi_{i,\vec{x}}](\vec{x})}{ 2 a^2 dx^2} + V^{(2)}(\phi_{1,\vec{x}},...,\phi_{N,\vec{x}})\Bigg).\\
\end{aligned}
\end{equation}
In the equations of motion corresponding to this Hamiltonian we will now assume that the perturbation part $\mathcal{H}^{\textrm{(2)}}$ is negligible compared to $\mathcal{H}^{\textrm{(0)}}$ and does not affect the dynamics of the scale factor or the homogeneous field equations. This way the background equations can be solved independently of the perturbations. 

With this reasoning we can write the Hamiltonian equations
\begin{equation} \label{eq:a-lin}
\begin{aligned}
a' &= \frac{\partial \mathcal{H}}{\partial p_a} = - \frac{p_a}{ 6 V_L m_{Pl}^2}\\
p_a' &= - \frac{\partial \mathcal{H}}{\partial a} = \sum_{i} V_L a^3 \Bigg( \frac{\pi_{\textrm{0},i}^2}{a^6} - 4 V(\phi_{\textrm{0},1},...,\phi_{\textrm{0},N})\Bigg).
\end{aligned}
\end{equation}
and for the homogeneous scalar field $i$
\begin{equation} \label{eq:phi-0}
\begin{aligned}
\phi_{\textrm{0},i}' &= \frac{\partial \mathcal{H}}{\partial (\pi_{\textrm{0},i})} = \frac{\pi_{\textrm{0},i}}{a^2} \\
\pi_{\textrm{0},i}' &= -\frac{\partial \mathcal{H}}{\partial (\phi_{\textrm{0},i})} = - a^4 \frac{\partial V }{\partial (\phi_{\textrm{0},i})}
\end{aligned}
\end{equation}
and the perturbation evolution is given by
\begin{equation} \label{eq:phi-1}
\begin{aligned}
\delta\phi_{i,\vec{z}}' &= \frac{\partial \mathcal{H}}{\partial (\delta\pi_{i,\vec{z}})} = \frac{\delta\pi_{i,\vec{z}}}{a^2} \\
\delta\pi_{i,\vec{z}}' &= -\frac{\partial \mathcal{H}}{\partial (\delta\phi_{i,\vec{z}})} =  a^2 \frac{D[\delta\phi_{i,\vec{z}}](\vec{z})}{ dx^2} - a^4 \frac{\partial V^{(2)} }{\partial (\delta\phi_{i,\vec{z}})}.\\
\end{aligned}
\end{equation}
Since these perturbation equations are linear we have chosen to evolve these in Fourier space.
The corresponding equations read
\begin{equation} \label{eq:Phik-1}
\begin{aligned}
\delta\Phi_{i,\vec{k}}' & = \frac{\delta\Pi_{i,\vec{k}}}{a^2} \\
\delta\Pi_{i,\vec{k}}' &=  \frac{-k_{\textrm{eff}}^{2} a^2 \delta\Phi_{i,\vec{k}}}{ dx^2} - a^4 \mathcal{F}\Big(\frac{\partial V^{(2)} }{\partial (\delta\phi_{i,\vec{z}})}\Big),\\
\end{aligned}
\end{equation}
where $-k_{\textrm{eff}}^{2}$ is the wave number corresponding to the discrete Laplacian operator $D[\phi](\vec{z})$,  $\mathcal{F}()$ is the discrete Fourier transform operator and the term inside the brackets can be easily read from equation (\ref{eq:V-2}). Note that inside the lattice there are multiple positions at which $k_{\textrm{eff}}^{2}$ terms get identical values and hence the perturbation equations of field $i$ are identical at these positions and any differences in their evolved values originate from different initial values. We use this observation to reduce the number of solved differential equations significantly.

\subsection{Symplectic integration}

After constructing the discretized Hamiltonian functions (\ref{eq:H2}) and (\ref{eq:H-lin}) the next task is to select a suitable integration method. Remembering that the right hand side of the Hamiltonian corresponds to the first Friedmann equation (\ref{eq:Fried-1}) a natural selection is an integrator that preserves the value of the Hamiltonian leading to a solution consistent with the Friedmann equations.

Symplectic integrators are a group of integrators that have this property and are therefore ideal for the task. A further 
simplification comes from the observation that the Hamiltonian functions (\ref{eq:H2}) and (\ref{eq:H-lin}) can be split into parts that can be integrated explicitly. By composing these sub-integrators suitably together higher order integrators of the whole system can be constructed. A good introduction to the subject of different splitting methods and geometric integrators can be found for example in \cite{McLachlan:2002}.

In general the equations of motion of a Hamiltonian system with canonical variable and momentum $z=(q,p)$ can be written in the form
\begin{equation} \label{eq:z}
z' = \{z,\mathcal{H}\} = \mathcal{D}_{\mathcal{H}}z,
\end{equation}
where we have introduced the operator $\mathcal{D}_{\mathcal{H}} = \{,\mathcal{H}\}$ and $\{,\}$ is the Poisson bracket.
The formal solution of this equation reads 
\begin{equation} \label{eq:z2}
z(\eta) = \exp(\eta \mathcal{D}_{\mathcal{H}}) z(0).
\end{equation}
Assuming that the Hamiltonian can now be split into two integrable parts $\mathcal{H}_i$, $i=1,2$ the task is to find a suitable composition of the operators $\exp(\eta \mathcal{D}_{\mathcal{H}_1})$ and $\exp(\eta \mathcal{D}_{\mathcal{H}_2})$ that approximates equation (\ref{eq:z2}) up to some order. More formally a set of real coefficients $c_{i}$ and $d_{i}$ need to be determined in the following equality
\begin{equation} \label{eq:z3}
\exp\Big(\eta (\mathcal{D}_{\mathcal{H}_1}+\mathcal{D}_{\mathcal{H}_2})\Big) = \prod_i^k \exp\Big(c_i \eta \mathcal{D}_{\mathcal{H}_1}\Big)\exp\Big(d_i \eta\mathcal{D}_{\mathcal{H}_2}\Big) + \mathcal{O}(\eta^{n+1}).
\end{equation}
This can be done for example with the help of the famous Baker-Campbell-Hausdorff formula. See \cite{Yoshida:1990} for further details.
For example a second order accurate symplectic integrator corresponds to values $c_{1}=1/2$, $d_{1}=1$ and $c_{2}=1/2$
and reads
\begin{equation} \label{eq:Phi-2}
\Phi^{(2)}(\eta) = \exp\left(\frac{\eta}{2}\mathcal{D}_{\mathcal{H}_1}\right) \exp\left( \eta\mathcal{D}_{\mathcal{H}_2}\right) \exp\left(\frac{\eta}{2}\mathcal{D}_{\mathcal{H}_1}\right).
\end{equation}


Second-order integrator however might not be precise enough for all applications and higher order methods are needed.
One way to proceed is to compose higher order integrators from a second order one by recursion \cite{Yoshida:1990}
\begin{equation} \label{eq:Phi-rec}
\Phi^{(2n+2)}(\eta) = \Phi^{(2n)}(z_1\eta)\Phi^{(2n)}(z_0\eta)\Phi^{(2n)}(z_1\eta),
\end{equation}
where the different coefficients equal
\begin{equation}
\begin{aligned}
z_0 & = -\frac{2^{\frac{1}{2n+1}}}{2-2^{\frac{1}{2(2n+1)}}} \\
z_1 & = \frac{1}{2-2^{\frac{1}{2n+1}}}.
\end{aligned}
\end{equation}
The downside of this method is that the number of times $\Phi_2(\eta)$ is evaluated by an integrator of order $2k$ is $3^{k-1}+1$
and hence grows exponentially. We have therefore used this method only at 4th order integrations.

A remedy to this problem \cite{Yoshida:1990} is to consider symmetric integrators composed of second-order symplectic maps
\begin{equation}
\Phi^{(n)}(\eta) = \Phi_{2}(w_s\eta)\Phi_{2}(w_{s-1}\eta) \cdots \Phi_{2}(w_2\eta)\Phi_{2}(w_1\eta),
\end{equation}
where the coefficients satisfy a symmetricity condition $w_{s+1-i} = w_{i}$ and $s$ is odd. This way the number of steps needed can be reduced dramatically: at 8th order it is possible to compose an integrator with 15 evaluations of $\Phi_{2}(\eta)$ compared to 28 iterations needed with the recursive method (\ref{eq:Phi-rec}). In Table 1 we have listed the values of $w_{i}$ for $i=1,...,(s+1)/2$ used in PyCOOL taken from \cite{Yoshida:1990,McLachlan:1995}. Note that the rest of the values follow from symmetry.

\begin{table}[h!b!p!]
\begin{center}
\begin{tabular}{|c|c|l|}
\hline
Order & Stages & Coefficients \\
\hline
\hline
 & & $w_1 = 0.78451361047755726382$  \\
 6 & 7 & $w_2 = 0.23557321335935813368$  \\
 & & $w_3 = -1.17767998417887100695$  \\
 & & $w_4 = 1 - 2(w_1+w_2+w_3)$  \\
 \hline
 & & $w_1 = 0.74167036435061295345$ \\
 8 & 15 & $w_2 = -0.40910082580003159400$  \\
 & & $w_3 = 0.19075471029623837995$  \\
 & & $w_4 = -0.57386247111608226666$  \\
 & & $w_5 = 0.29906418130365592384$  \\
 & & $w_6 = 0.33462491824529818378$  \\
 & & $w_7 = 0.31529309239676659663$  \\
 & & $w_8 = 1 - 2\sum_{i=1}^7 w_i$  \\
\hline
\end{tabular}
\label{table1}
\caption{List of the used integrator coefficients in PyCOOL.}
\end{center}
\end{table}

A suitable splitting of the Hamiltonian (\ref{eq:H2}) derived previously is
\begin{equation} \label{eq:split-H}
\begin{aligned}
\mathcal{H}_1 & = -\frac{ p_a^2 }{12 V_L m_{Pl}^2}, \\
\mathcal{H}_2 & = \sum_{i,\vec{x}} a^4 \left( \frac{\pi_{i,\vec{x}}^2}{2 a^6} \right) + a^4\left(\frac{ V_L \rho_{\gamma,0}}{a^4} + \frac{ V_L \rho_{m,0}}{a^3}\right),\\
\mathcal{H}_3 & = \sum_{i,\vec{x}} a^4 \left( - \frac{\phi_{i,\vec{x}}D[\phi_{i,\vec{x}}](\vec{x})}{ 2 a^2 dx^2} + V(\phi_{1,\vec{x}},...,\phi_{N,\vec{x}})\right)\\
\end{aligned}
\end{equation}
since now the scale factor is evolved by $\mathcal{H}_1$, scalar fields by the second Hamiltonian and the canonical momentums of the fields in the last part. The canonical momentum of scale factor is evolved in $\mathcal{H}_2$ and $\mathcal{H}_3$.
The Hamiltonian equations related to these can be integrated explicitly to give flows
\begin{equation} \label{eq:split-Phi}
\begin{aligned}
\Phi_{\mathcal{H}_1}(d\eta) : \Bigg(a, p_a, \phi_{i,\vec{x}}, \pi_{i,\vec{x}} \Bigg) & \mapsto \Bigg(a - \frac{p_a}{ 6 V_L m_{Pl}^2} d\eta, p_a, \phi_{i,\vec{x}}, \pi_{i,\vec{x}} \Bigg), \\
\Phi_{\mathcal{H}_2}(d\eta) : \Bigg(a, p_a, \phi_{i,\vec{x}}, \pi_{i,\vec{x}} \Bigg) & \mapsto  \Bigg(a, p_a + \Bigg[ \sum_{i,\vec{x}} a^3 \Bigg( \frac{\pi_{i,\vec{x}}^2}{a^6}\Bigg) - V_L \rho_{m,0} \Bigg] d\eta , \phi_{i,\vec{x}} +  \frac{\pi_{i,\vec{x}}}{a^2} d \eta , \pi_{i,\vec{x}}\Bigg)\\
\Phi_{\mathcal{H}_3}(d\eta) : \Bigg(a, p_a, \phi_{i,\vec{x}}, \pi_{i,\vec{x}} \Bigg) & \mapsto \Bigg(a, p_a  + \Bigg[ \sum_{i,\vec{x}} a^3 \Bigg(\frac{\phi_{i,\vec{x}}D[\phi_{i,\vec{x}}](\vec{x})}{ a^2 dx^2} - 4 V(\phi_{1,\vec{x}},...,\phi_{N,\vec{x}})\Bigg)\Bigg] d\eta , \phi_{i,\vec{x}},\\
& \qquad \qquad  \pi_{i,\vec{x}} + \Bigg[a^2 \frac{D[\phi_{i,\vec{x}}](\vec{x})}{ dx^2} - a^4 \frac{\partial V }{\partial (\phi_{i,\vec{x}})}\Bigg] d\eta \Bigg)\\
\end{aligned}
\end{equation}
By writing $\mathcal{H} = \mathcal{H}_1 + \tilde{\mathcal{H}}_2 = \mathcal{H}_1 + \mathcal{H}_2 + \mathcal{H}_3$ and by performing the splitting used in equation (\ref{eq:Phi-2}) twice a second order integrator scheme of the scalar field system reads
\begin{equation}
\Phi^{(2)}_{\mathcal{H}}(d\eta) = \Phi_{\mathcal{H}_1}\left(\frac{d\eta}{2}\right)\Phi_{\mathcal{H}_2}\left(\frac{d\eta}{2}\right) \Phi_{\mathcal{H}_3}(d\eta) \Phi_{\mathcal{H}_2}\left(\frac{d\eta}{2}\right) \Phi_{\mathcal{H}_1}\left(\frac{d\eta}{2}\right).
\end{equation}
Higher order integrators can now be easily constructed from this expression with the methods presented above.



A splitting of the linearized Hamiltonian (\ref{eq:H-lin}) is similar to the non-linear case. However now the perturbation part gives two additional Hamiltonians:
\begin{equation} \label{eq:split-H-lin}
\begin{aligned}
\mathcal{H}_1 & = -\frac{ p_a^2 }{12 V_L m_{Pl}^2}, \\
\mathcal{H}_2 & = \sum_{i} a^4 V_L \left( \frac{\pi_{\textrm{0},i}^2}{2 a^6} \right) + a^4\left(\frac{ V_L \rho_{\gamma,0}}{a^4} + \frac{ V_L \rho_{m,0}}{a^3}\right),\\
\mathcal{H}_3 & = \sum_{i} a^4 V_L \left( V^{(0)}(\phi_{\textrm{0},1},...,\phi_{\textrm{0},N})\right)\\
\mathcal{H}_4 & = \sum_{i,\vec{x}} a^4 \left( \frac{\delta\pi_{i,\vec{x}}^2}{2 a^6} \right),\\
\mathcal{H}_5 & = \sum_{i,\vec{x}} a^4 \left( - \frac{\delta\phi_{i,\vec{x}}D[\delta\phi_{i,\vec{x}}](\vec{x})}{ 2 a^2 dx^2} + V^ {(2)}(\phi_{1,\vec{x}},...,\phi_{N,\vec{x}})\right).\\
\end{aligned}
\end{equation}
By writing $\mathcal{H} = \mathcal{H}_1 + \tilde{\mathcal{H}}_2 = \dots = \mathcal{H}_1 + \mathcal{H}_2 + \mathcal{H}_3 + \mathcal{H}_4 + \mathcal{H}_5$ and by using equation (\ref{eq:Phi-2}) a second order integrator scheme of the perturbed system reads
\begin{equation}
\begin{aligned}
\Phi^{(2)}_{\mathcal{H}}(d\eta) = & \Phi_{\mathcal{H}_1}\left(\frac{d\eta}{2}\right)\Phi_{\mathcal{H}_2}\left(\frac{d\eta}{2}\right) \Phi_{\mathcal{H}_3}\left(\frac{d\eta}{2}\right) \Phi_{\mathcal{H}_4}\left(\frac{d\eta}{2}\right) \Phi_{\mathcal{H}_5}(d\eta) \times \\
& \Phi_{\mathcal{H}_4}\left(\frac{d\eta}{2}\right) \Phi_{\mathcal{H}_3}\left(\frac{d\eta}{2}\right) \Phi_{\mathcal{H}_2}\left(\frac{d\eta}{2}\right) \Phi_{\mathcal{H}_1}\left(\frac{d\eta}{2}\right).\\
\end{aligned}
\end{equation}
The order of this integrator can be verified for example with Huang's Python script that is available at 
\href{http://www.cita.utoronto.ca/~zqhuang/work/symp6.py}{http://www.cita.utoronto.ca/~zqhuang/work/symp6.py}.

When integrating the perturbation equations we use a similar method that was given in \cite{Chambers:2009ki} to reduce the number of the integrated differential equations.
In general the solution of a linear differential equation group can be written as
\begin{equation} \label{eq:pert-evo}
\begin{aligned}
\begin{pmatrix}
\phi_1(\tau,k_\textrm{eff})  \\
\pi_1(\tau,k_\textrm{eff})  \\
\vdots    \\
\phi_N(\tau,k_\textrm{eff})  \\
\pi_N(\tau,k_\textrm{eff})  \\
\end{pmatrix}
= M(\tau,k_\textrm{eff})
\begin{pmatrix}
\phi_1(0,k_\textrm{eff})  \\
\pi_1(0,k_\textrm{eff})  \\
\vdots    \\
\phi_N(0,k_\textrm{eff})  \\
\pi_N(0,k_\textrm{eff})  \\
\end{pmatrix}
\end{aligned}
\end{equation}
where $M(\tau,k_\textrm{eff})$ is an $n\times n$ matrix and we have written explicitly the effective mode dependence of the equations. We now use this equation to evolve the field perturbations with the following method: by remembering that in the lattice the effective wave number $k_\textrm{eff}^2$ will get equal values at a number of different locations we will first calculate the distinct values of $k_\textrm{eff}^2$ and then evolve the growth functions of a given field related to the modes $M(\tau,k_\textrm{eff}^2) \mathbf{1}_{i}$, where $\mathbf{1}_{i}^\mathrm{T} = (0,\dots,1,\dots,0)$, \ie a vector with one at position $i$. We then write the right hand side of equation (\ref{eq:pert-evo}) as a linear combination of the different initial values $\phi_i(0,k_\textrm{eff})$ and $\pi_i(0,k_\textrm{eff})$ and use the solved growth functions to evolve the Fourier modes of the perturbations.

\subsection{PyCUDA implementation}

The symplectic integrators presented above have to be naturally also implemented in some computer language. We have chosen to use a combination of Python and Nvidia's CUDA GPGPU language to make a program that is easy to adapt to different models of preheating but is also computationally very fast. The use of Python generally leads to some overhead compared to a pure C-language program and hence a slower program but the easy of use of the program more than makes up this deficiency.

The program uses object-oriented programming extensively: the lattice, the potential function, the simulation and the fields are defined as python objects whereas the symplectic integration functions are implemented as subroutines of a python object that holds all the CUDA functions used in the integration steps.
The program also uses extensively textual templating to write the necessary CUDA codes.
After the end user has written a model file that defines the necessary fields, potential functions and the initial values the program will read these variables and values and write the necessary CUDA codes from predefined template files.
This means that in ideal situations the end user only has to write the potential functions of the fields as a list of strings that PyCOOL then implements in suitable CUDA form with Python's SymPy package and custom string formatting codes written for the program.
The final CUDA kernels are written for debugging and verification purposes into a separate folder where they can be easily inspected for any errors.

The initialization of the different fields is done as in Defrost with a convolution based method. The details of this method can be found in \cite{Frolov:2008hy}. These routines are included in a separate file that can be easily modified to include additional initialization algorithms.

The implementation of the non-linear symplectic integrator (\ref{eq:split-Phi}) on GPU is similar to the one used in CUDAEASY \cite{Sainio:2009hm} with some major upgrades. We will use similar computational labor division as in CUDAEASY where the scalar fields were evolved on the GPU whereas the scale factor evolution was calculated with the CPU. The main differences come from the use of the very precise symplectic algorithms compared to a second order leap-frog algorithm and the use of double precision calculations.
We have also optimized the CUDA kernels to use fewer registers compared to CUDAEASY which means that in theory more thread blocks can be calculated simultaneously. However the use of double precision arithmetics roughly doubles the number of the used registers which somewhat negates these improvements.

We will briefly go through the steps involved in the GPU evolution. Further details can be found in the CUDAEASY paper \cite{Sainio:2009hm} where the used CUDA terms were also defined. CUDA implementation of 3D stencils needed in the $\mathcal{H}_3$ part was presented in detail in \cite{Micikevicius:2009}. We will follow a similar method that uses shared memory to calculate the Laplacian operators. Because of the limited amount of shared memory of the GPU the 3 dimensional lattice has to be divided into smaller pieces. To accomplish this we have sliced the lattice along the $z$-axis into smaller tiles and advance these tiles in $z$-direction. The computations within these tiles are done by CUDA thread blocks. This means that a single thread of a thread block will advance all the scalar fields of a column with constant $x$- and $y$-coordinates. Since the outer threads of a thread block need the values of scalar fields that are in a different block and since different blocks can only communicate through global memory the role of the outer threads of a thread block is only to load data into shared memory and the scalar field computations are done by the inner threads.
The computation starts from the bottom of the lattice \ie at $z=0$ and proceeds to the top.

The integration of $\phi_{i,\vec{x}}$ does not include any stencil operations that makes this step computationally much easier. Summations over the lattice that are needed when updating $p_a$ are performed also in the GPU and are done simultaneously with the evolution steps of $\phi_{i,\vec{x}}$ and $\pi_{i,\vec{x}}$.
The energy density and pressure of the scalar fields \ie equations (\ref{eq:rho-pres}) are also calculated on the GPU. 
However since the symplectic evolution equations (\ref{eq:split-Phi}) do not explicitly depend on these variables they are not calculated at every step. This way the number of times the computationally demanding $G[\phi_{i,\vec{x}}]$ terms given in equation (\ref{eq:G}) need to be calculated can be reduced significantly. How frequently to calculate the energy densities and to write them to a file is decided by the user.

The linearized code is somewhat different and we will briefly explain also the main steps involved. The main difference is that we now evolve all of the equations on the GPU. The used CUDA grid is chosen to be such that every thread calculates the evolution of all of the growth functions $M(\tau,k_\textrm{eff}^2) \mathbf{1}_{i}$ that have the same value of $k_\textrm{eff}^2$, \ie there is a CUDA thread per a distinct value of $k_\textrm{eff}^2$. The evolution of all the background variables is done at the CUDA block level meaning that the scale factor and the canonical momentum $p_a$ are evolved by the first thread of a block. These values are stored in the shared memory which is visible to every thread in a block. Compared to the non-linear code where scale factor was evolved by the CPU this way the memory transfers between GPU and the CPU can be eliminated which results in a further speedup of the code. This way different thread blocks don't have to be synchronous meaning that no thread block idles during the calculation steps.
After the growth functions have been integrated a given number of steps forward we use a CUDA kernel to read the growth functions related to a position in Fourier space in the lattice and then evolve the perturbation mode with equation (\ref{eq:pert-evo}). Overall this method leads to roughly 15 times faster code compared to a naive version where every perturbation field mode is evolved independently.

Note that the linearization part has been tested only with polynomial potential models and in general the use of the non-linear integrator is preferred. Hopefully in future releases the linearized code is expanded further and improved.

The results of the program are written intermittently into a SILO file that can be easily plotted with VisIt program provided by LLNL. After the simulation of dynamics is done we have included a post processing phase which calculates a number of different spectra. These include field, number density and energy density spectra that are used in LATTICEEASY program \cite{Felder:2000} and the field spectrum algorithm used in Defrost \cite{Frolov:2008hy}. These post processing functions have been coded in Cython in order to make these computations much faster compared to ordinary Python implementation. The results are written as curves in the SILO file which can be easily plotted in the VisIt program.

It is also possible to perform non-gaussianity and curvature perturbation related calculations with PyCOOL. This is done by using the $\Delta N$ formalism and the separate universe approximation \cite{Chambers:2007se,Chambers:2008gu,Bond:2009xx} when solving the evolution of the system with different initial values. We will not however apply this functionality in this paper as it will be done in a follow-up paper
where we study the curvaton preheating in closer detail.

\section{Numerical results}

We have tested the functionality of the program with three different scalar field models. We will use units where the reduced Planck mass $m_{\textrm{pl}}=(8\pi G)^{-1/2}$ is set to one in all of the simulations.
Note also that only the model file that defines the necessary fields, potential functions and the initial values changes from one simulation to other and the actual evolution code stays unchanged.

We use the absolute value of the relative residual curvature
\begin{equation} \label{eq:res-curv}
\frac{K}{a^2 H^2} = \Big|\frac{8\pi G \langle \rho \rangle }{3 H^2} - 1\Big|
\end{equation}
to measure the conservation of Hamiltonian in all of the examples.

We have included plots of some of the spectra and post-processing data generated by the program in this section. For further examples of the output generated by the program see the home page \href{http://www.physics.utu.fi/tiedostot/theory/particlecosmology/pycool/}{http://www.physics.utu.fi/tiedostot/theory/particlecosmology/pycool/}.

\subsection{Chaotic inflation model}

We have first tested the program by solving the evolution of scalar fields in the simple chaotic inflation model
which has become a standard test scenario studied with LATTICEEASY \cite{Podolsky:2005bw}, DEFROST \cite{Frolov:2008hy},  CUDAEASY \cite{Sainio:2009hm} and with PSpectRe \cite{Easther:2010qz}.
The potential in this case is
\begin{equation}
V(\phi,\psi) = \frac{1}{2}m^2\phi^2 + \frac{1}{2}g^2\phi^2\psi^2,
\end{equation}
where $\phi$ is the inflaton and $\psi$ the decay product.

The initial values have been set to coincide with the values given in \cite{Podolsky:2005bw,Frolov:2008hy}. This means that homogeneous value of the inflaton is set to $\phi \simeq 1.009343$ and the decay field $\psi = 0$, the mass of the inflaton equals $5 \cdot 10^{-6} m_{\textrm{pl}}$, the Hubble parameter $H \simeq 0.50467m$, and the value of the coupling constant $g=100 m$. We have evolved the system with conformal time step $d\eta = 0.0001/m$ and lattice size $256^3$ until $t_{\textrm{phys}} = 250/m$.

We have done the simulations with two different systems for comparison. The more expensive computer has a hexa core AMD processor, 16 GB of memory and an NVIDIA Tesla C2050 compute card whereas the much cheaper consumer computer has a quad core processor and an NVIDIA GTX 470 graphics card. Since NVIDIA has limited the double precision speed of the Fermi consumer graphics cards to one quarter of the Tesla series we can also test what kind of difference this makes in the simulation time. We use thread blocks that contain $324 = (18^2)$ threads of which computations are done by the inner 256 threads. This configuration was found to be the fastest of 64, 128 or 256 inner threads per block. The CUDA grid size is based on the size of lattice and for a $256^3$ lattice we use $256 (=16^2)$ thread blocks. Note that these simulations were run with identical seed values in the random number generators meaning that the initial random field values were also identical.

\begin{figure}[h]
\subfigure[]{\label{fig1a}\includegraphics*[width=0.45\columnwidth]{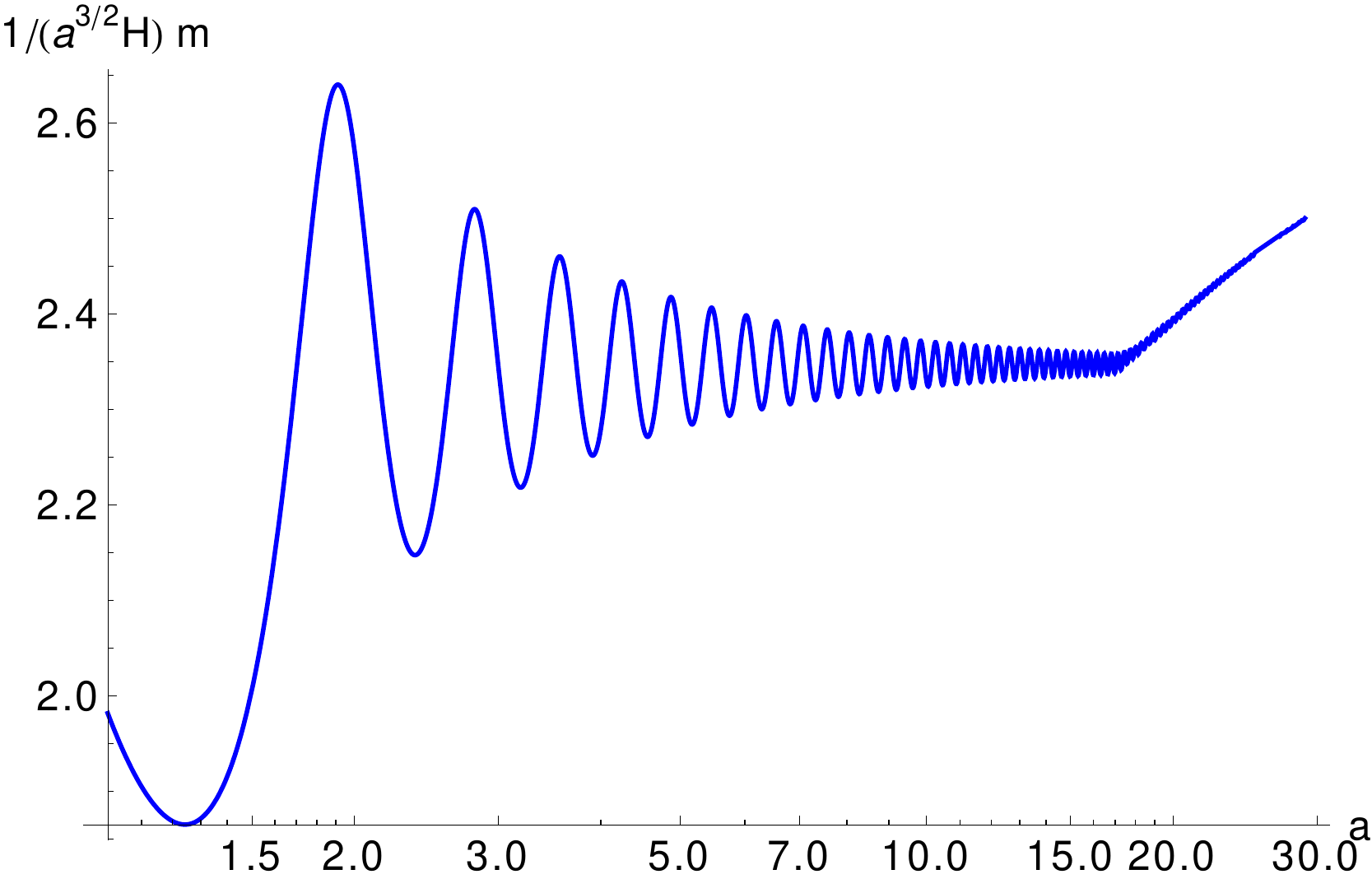}}
\quad
\subfigure[]{\label{fig1b}\includegraphics[width=0.45\columnwidth]{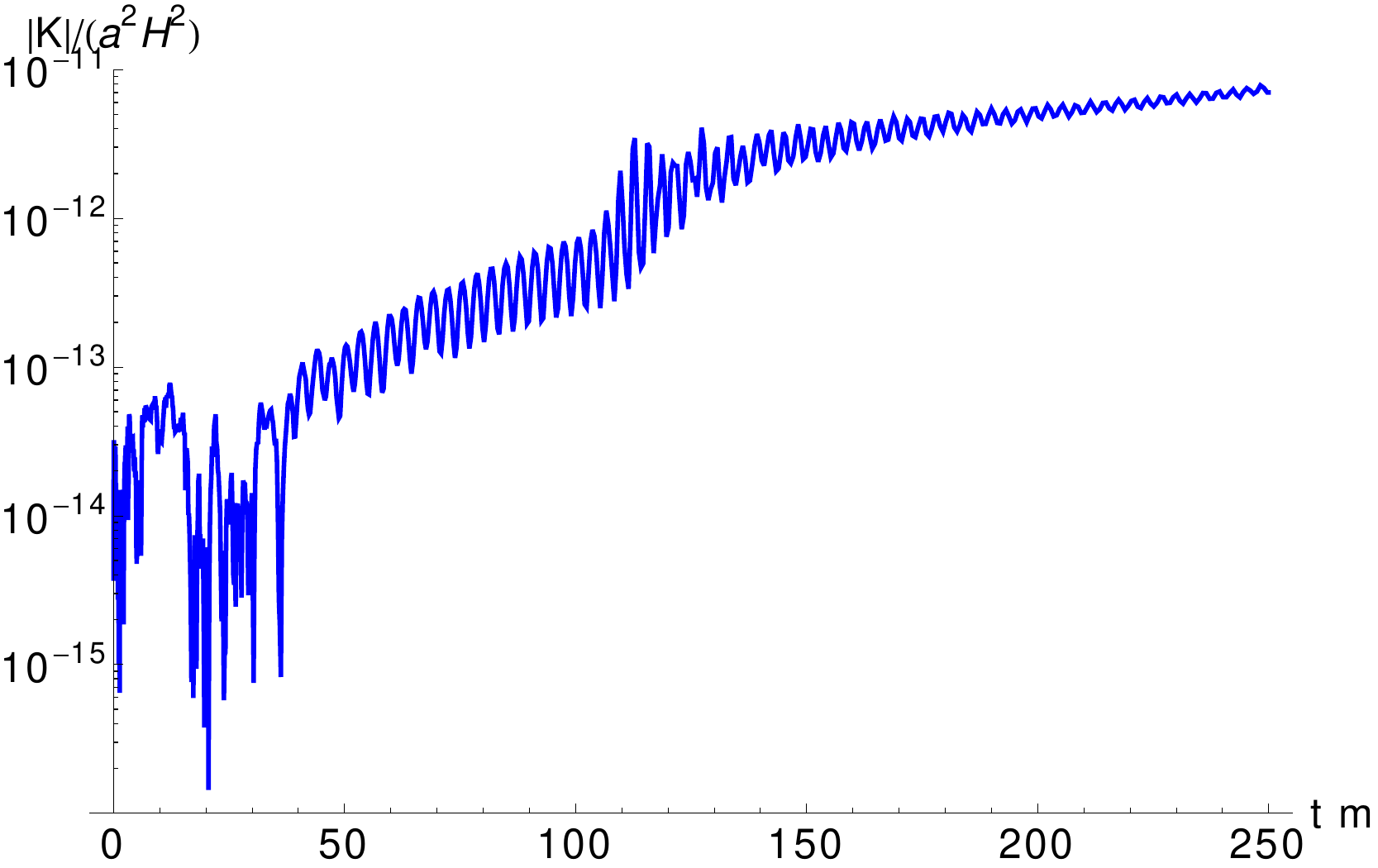}}
\subfigure[]{\label{fig1c}\includegraphics*[width=0.45\columnwidth]{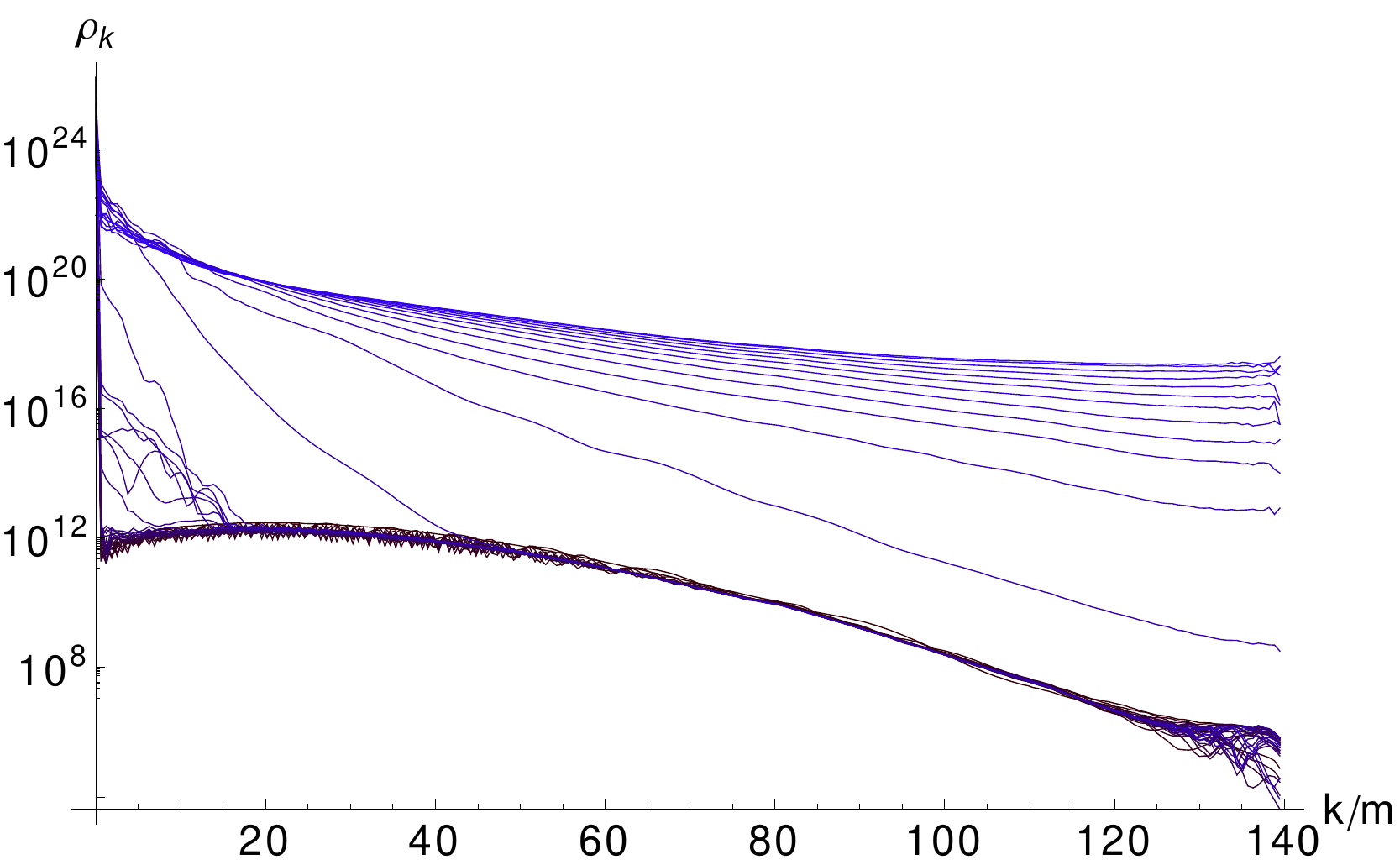}}
\quad
\subfigure[]{\label{fig1d}\includegraphics[width=0.45\columnwidth]{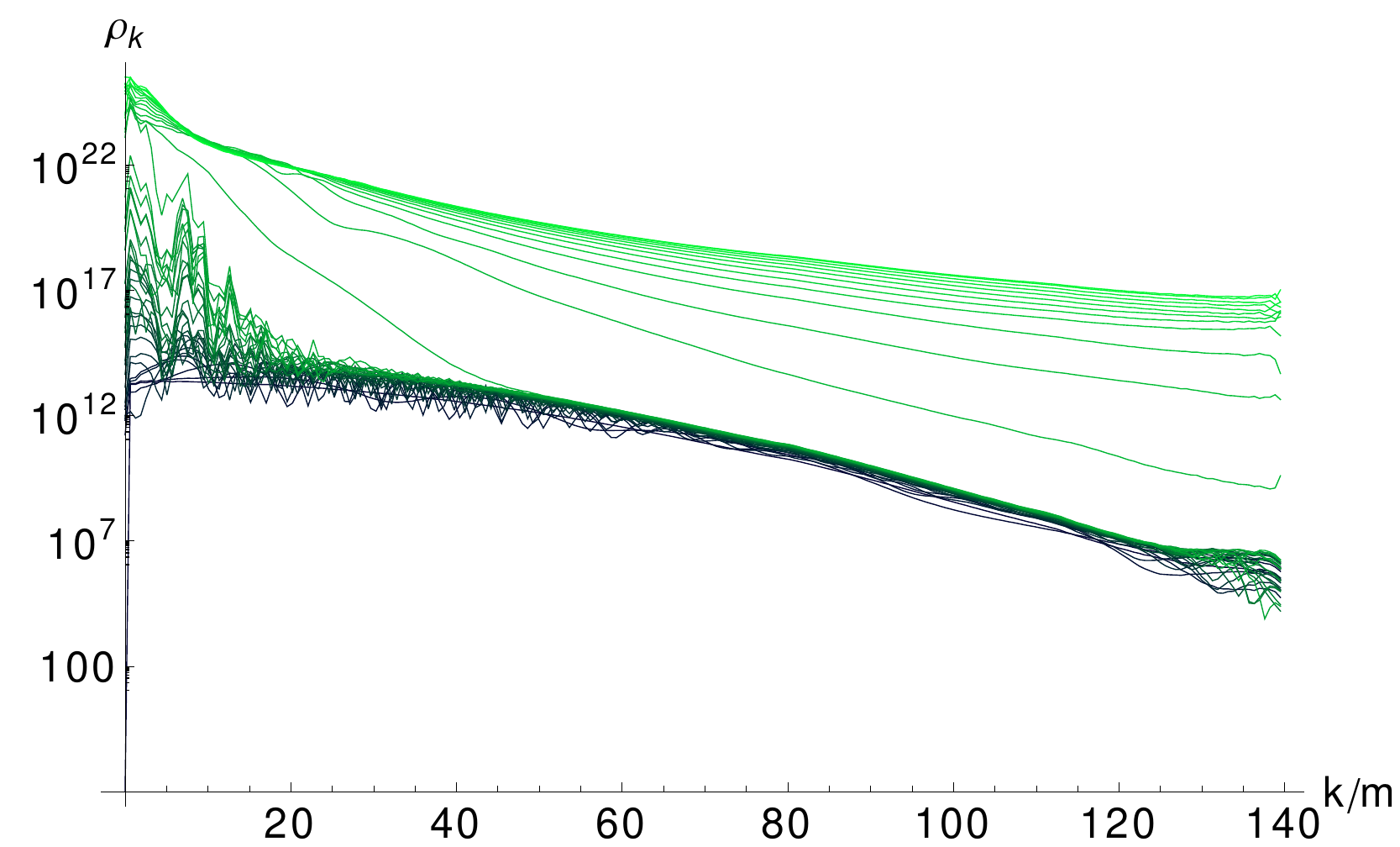}}
\caption{(a) The evolution $1/(a^{3/2}H)$ as a function of the scale parameter in the chaotic potential model.
(b) The absolute value of the relative residual curvature (\ref{eq:res-curv}) as a function of physical time in units of $1/m$. The energy density spectrum of (c) $\phi$ field and (d) $\psi$ field. Note that in these spectra the color evolves with time and the lightest curve (at the top) is calculated at $t_\textrm{phys}=250/m$ whereas darkest one (at the bottom) are evaluated at $t_\textrm{phys}=0/m$. Please see the online version of this article for color figures.}
\end{figure}

The numerical results for fourth order integrator are shown in figure \ref{fig1a} where we have plotted $a^{-3/2}H^{-1}$ as a function of the scale parameter. Note that in a matter dominated universe ($H^2 \propto a^{-3}$) this should be a constant. In figure \ref{fig1b} is the plot of the absolute value of the relative residual curvature (\ref{eq:res-curv}). Compared to the accuracy of Defrost and PSpectRe presented in \cite{Easther:2010qz} PyCOOL is more than 5 orders of magnitude more precise. Note however that by integrating in conformal time the step sizes increase in physical time as the scale factor increases. This is also what most likely causes the numerical error to increase in time. By using an adaptive conformal step with constant increases in physical time the numerical error will stop to increase. This will however lead to roughly 20 \% longer simulation times. Another possibility would be to solve the evolution of this model in physical time as is done usually \cite{Podolsky:2005bw,Frolov:2008hy}. We intend to add this possibility in a future version of PyCOOL.

We have included the energy density spectra of the scalar fields in figures \ref{fig1c} and \ref{fig1d}. In these figures it can be seen that the preheating process excites modes of the $\psi$ field with $k/m < 20$. These will eventually couple with the inflaton field leading to the phase of non-linear of dynamics of the system \cite{Podolsky:2005bw}.

The total computation for one step when using the fourth order integrator is on average 0.132 seconds when using a Tesla card which corresponds to a total running time of roughly eight hours for one simulation with $256^3$ elements. The GTX 470 card is 
roughly 15 \% slower integrating one step in 0.152 seconds. However by remembering that the double precision speed of the Tesla card is four times that of GTX 470 this difference can be seen to be minimal. When taking the rather high price of the Tesla cards also into account the consumer cards seem to offer a better choice for simple lattice calculations even at double precision accuracy.

When comparing these results to LATTICEEASY, Defrost and it should be remembered that we are using a fourth order accurate integrator. Lowering the order of the integrator to two which is also used in the aforementioned programs would lead to roughly three times faster computations but this would naturally lead to a lower accuracy of the simulations. A similar simulation as was used in Defrost with constant physical time steps and a second order integrator would take roughly three hours. For comparison a Defrost simulation execution speed is 0.32 s per step on quad core Xeon machine meaning that when using equal integrators PyCOOL is roughly six times faster. We also tested the speed of PSpectre on the faster test machine with a hexa core AMD processor. A test run showed that a $256^3$ simulation takes roughly 5.1 seconds per step meaning that PyCOOL is roughly 115 times faster when using integrators of equal order.

\subsection{Oscillons}

Oscillons are pseudo-stable, non-topological solitons that can form in the post-inflationary universe \cite{Gleiser:1993pt,Copeland:1995fq} . These have been studied numerically lately in \cite{Amin:2010dc} for one field case and in \cite{Gleiser:2011xj} for two field oscillons. The generation of these oscillons leads to a matter dominated phase after resonance which differs from the relativistic equation of state observed in the chaotic inflation after preheating.

We will simulate the oscillon production with a model that was used in \cite{Amin:2010dc}.
The potential function is set to be
\begin{equation}
V(\varphi) = \frac{1}{2}m \varphi^2 - \frac{\lambda}{4}\varphi^{4} + \frac{g^2}{6 m^2}\varphi^{6}
\end{equation}
where $\lambda > 0$ and $(\lambda/g)^2 \ll 1$ are necessary conditions for so called flat-top oscillons \cite{Amin:2010dc}. Note that this potential is assumed to be valid only in the post-inflationary universe. For the parameters and initial values we use the following values: $m = 5 \cdot 10^{-6} m_{\textrm{pl}}$, $\lambda = 2.8125\cdot 10^{-6}$, $g^2 = \lambda^ 2/0.1$. We also set the homogeneous value of the oscillon field to be $\varphi_0 = \sqrt{(3 \lambda )/(5 g^2) } \, m$ and the canonical momentum $\pi_0 = 10^{-16}m$. The lattice side length was set to $L = 400/m$ and  $d\eta = 0.005/m$. We will use a lattice of $128^3$ points whereas the simulations in \cite{Amin:2010dc} were done with at least $256^3$ points. Note that an additional post processing  function was used in \cite{Amin:2010dc} to detect the the oscillons from the data. Implementation of this function in PyCOOL is beyond the scope of this paper but could be included in a future version of the code.

\begin{figure}[h]
\subfigure[]{\label{fig2a}\includegraphics[width=0.45\columnwidth]{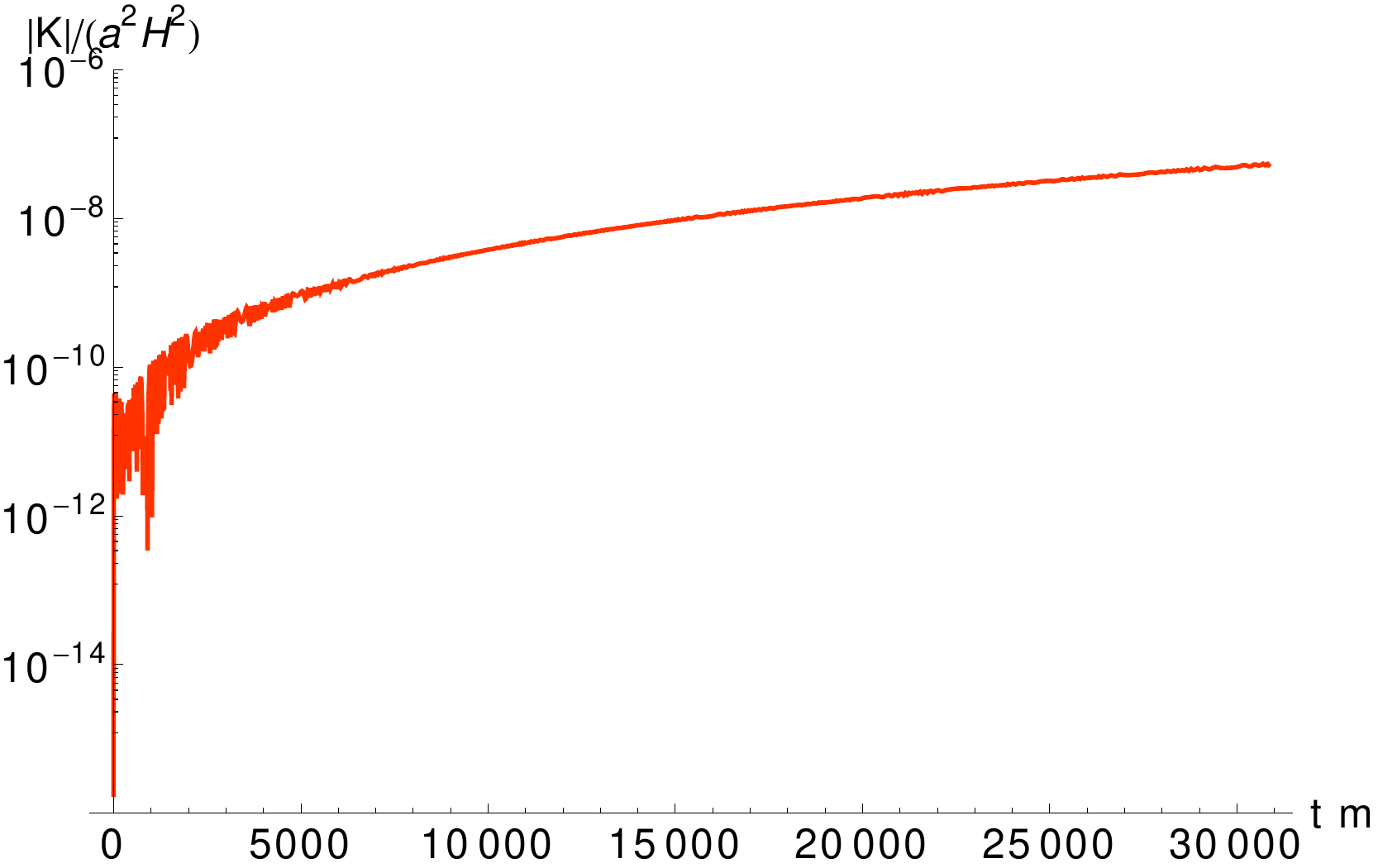}}
\quad
\subfigure[]{\label{fig2b}\includegraphics[width=0.45\columnwidth]{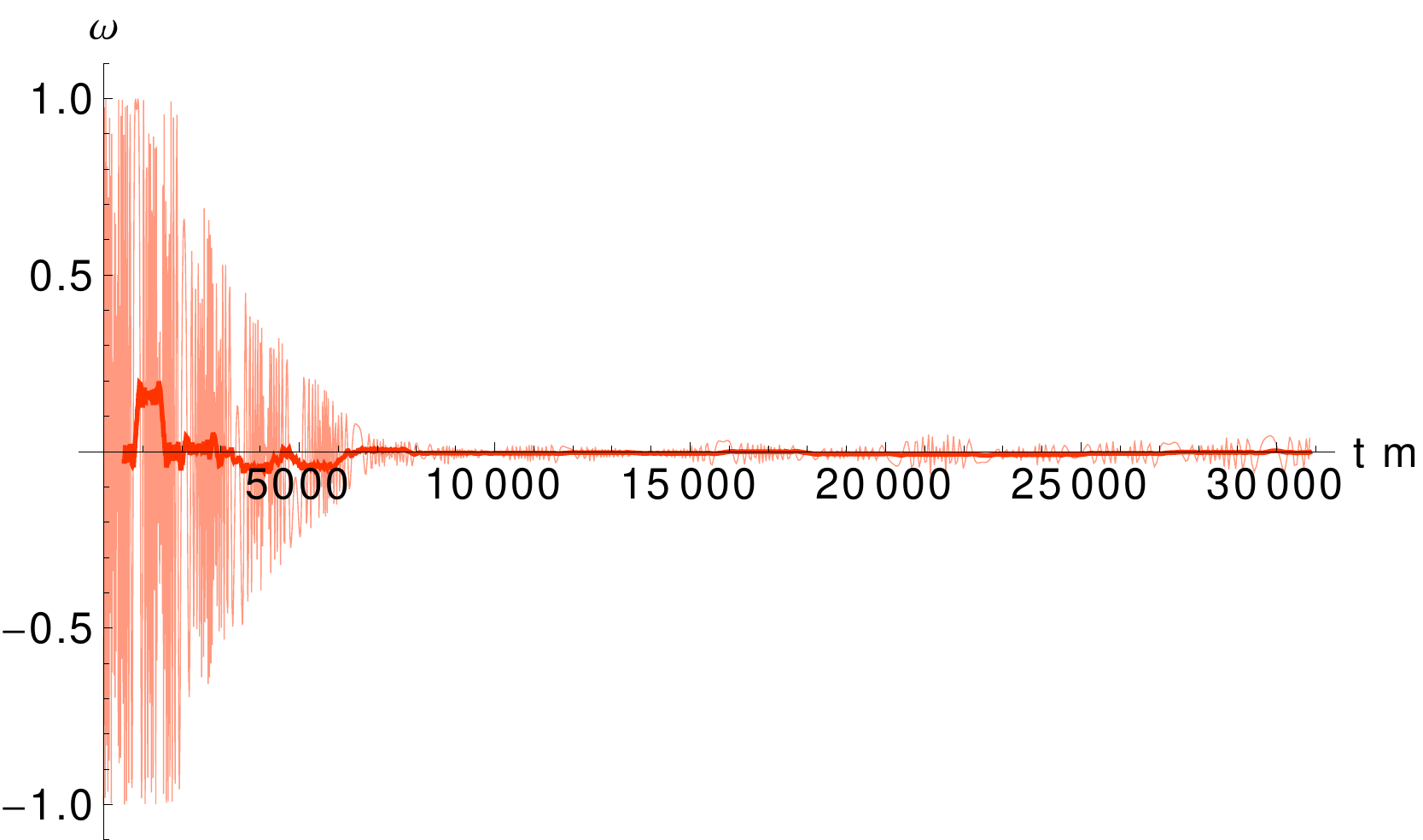}}
\subfigure[]{\label{fig2c}\includegraphics*[width=0.45\columnwidth]{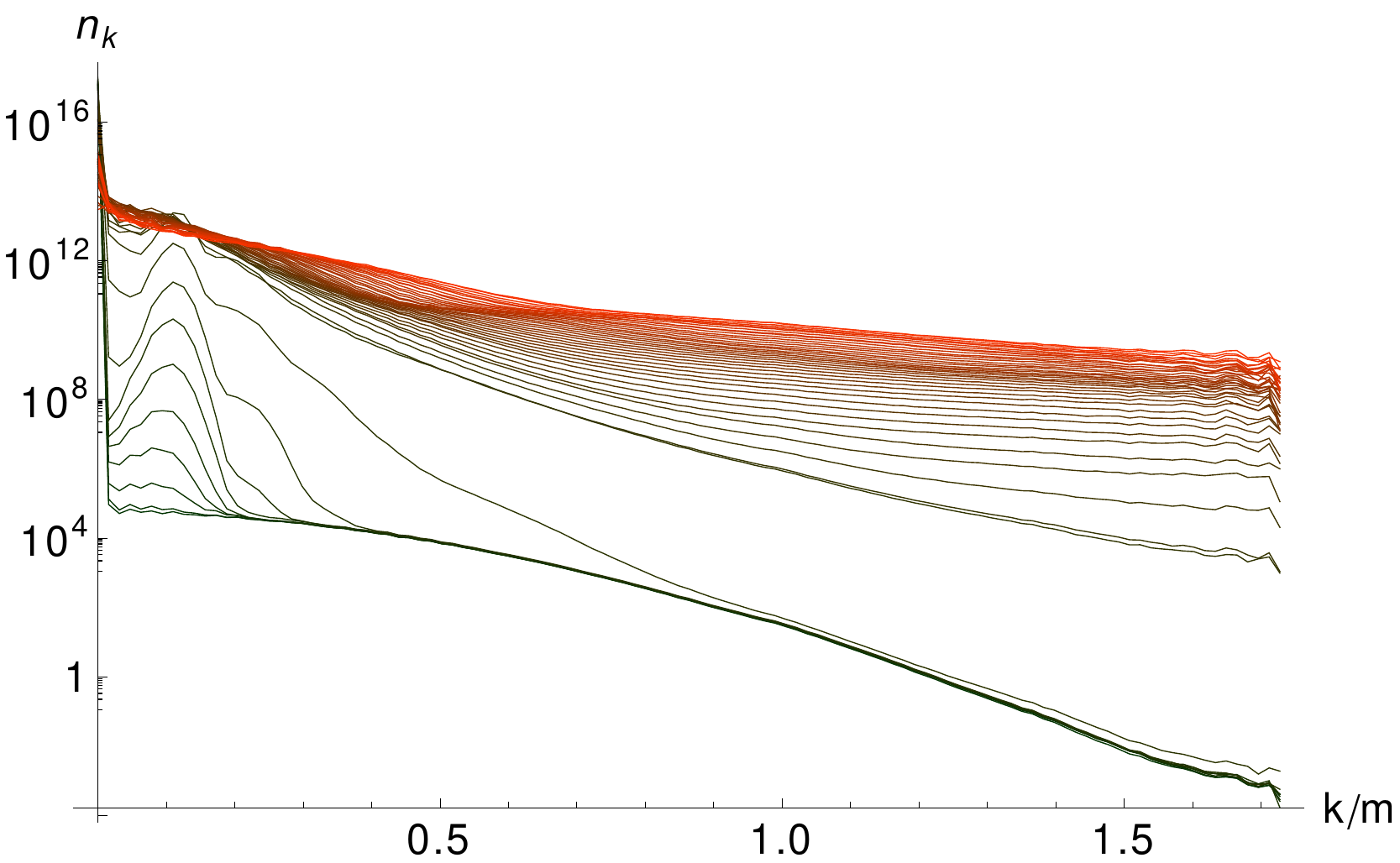}}
\quad
\subfigure[]{\label{fig2d}\includegraphics*[width=0.45\columnwidth]{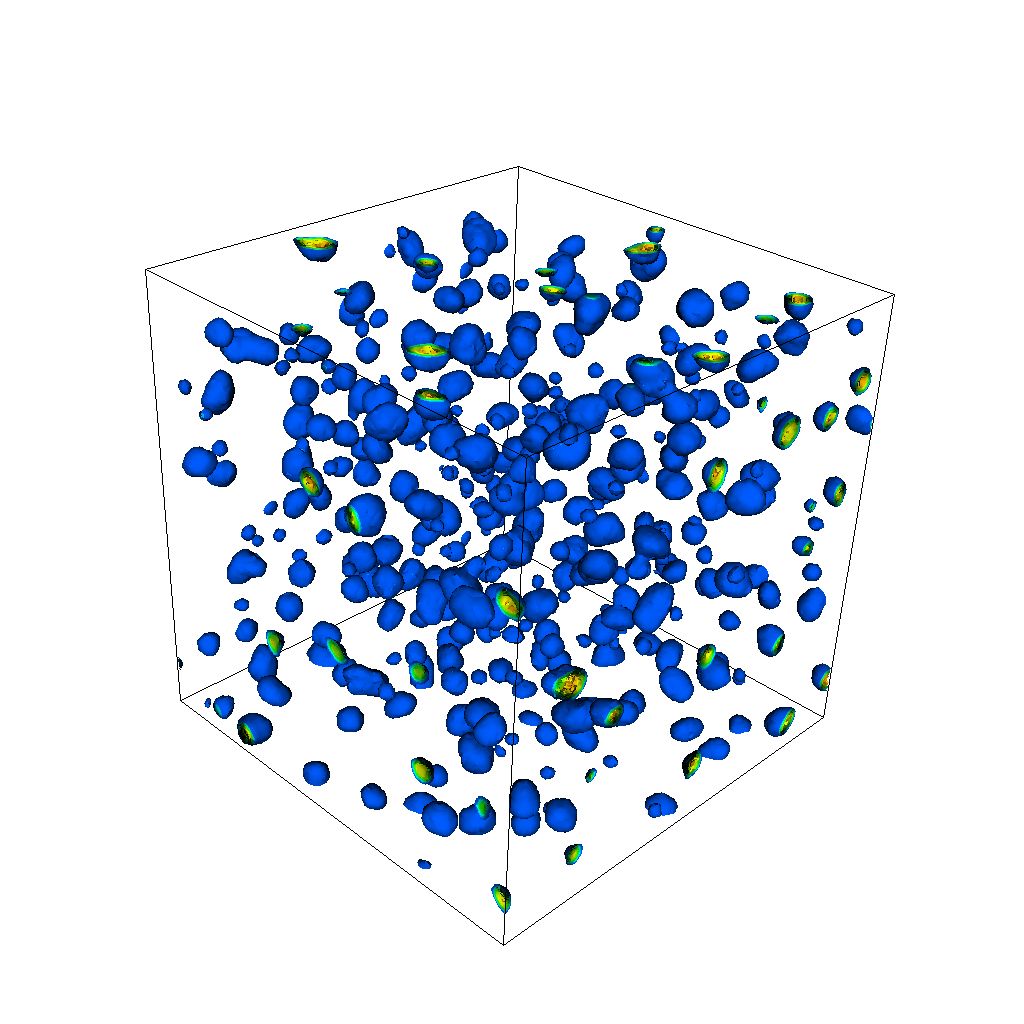}}
\caption{(a) The absolute value of the relative residual curvature (\ref{eq:res-curv}) as a function of physical time in units of $1/m$. (b) The equations of state of the field and a moving average of 50 steps over the data points used. The equation of state is clearly close to a non-relativistic matter.
The number density spectrum of (c) $\varphi$ field. Note that the color evolves with time and the lightest curve (at the top) is calculated at $t_\textrm{phys}=250/m$ whereas darkest one (at the bottom) are evaluated at $t_\textrm{phys}=0/m$.
In (d) we have the volume plot with isosurfaces of the energy density of the system calculated at $t_\textrm{phys} = 9821.88/m$. Please see the online version of this article for color figures.}
\end{figure}

We have plotted the evolution of $a^{-3/2}H^{-1}$ term as a function of the scale parameter in figure \ref{fig2a}. In figure \ref{fig2b} we the plot of the absolute value of the relative residual curvature (\ref{eq:res-curv})
which use again as a proxy for numerical accuracy. Compared to the results of chaotic potential simulations the algorithm is now less precise. We are however using a smaller lattice which will cause some of this error and a larger time step which is naturally another source for the observed inaccuracy.

We have also plotted the number density spectrum of oscillons for illustration in figure \ref{fig2c} with different shades of orange. The darkest curves correspond to initial values whereas the brightest curves are calculated close to the end of the simulation.
From the spectrum it can be seen that the fragmentation of the oscillon field happens initially close to $k \sim 0.1/m$ band in the momentum space. However as the system evolves other modes are excited as well and the peak close to $k \sim 0.1/m$ vanishes.
We have also included a snapshot of the energy density of the system in figure \ref{fig2d} at $t_\textrm{phys} = 9821.88/m$. The system is clearly seen to fragment into blobs of scalar matter.

These simulations were done with the Tesla card in 5 hours and 55 minutes leading to an average time of $0.0109$ seconds per one step. The same conclusions apply as previously: lowering the order of the integrator to two leads to three times faster simulations whereas increasing the size of the side of the lattice by two leads to eight times longer simulation times.

\subsection{Curvaton model with Q-balls}

The curvaton mechanism is a well studied alternative to the usual inflation scenario \cite{Mazumdar:2010sa}. The main idea is that the curvaton is a light scalar field that stays sub-dominant during the inflation process and does not contribute to the expansion of the space. 
After the inflation has ended the curvaton will eventually become massive compared to the Hubble rate $H$ and it will start to oscillate and eventually decay into relativistic particles.

Supersymmetric versions of the curvaton scenario have been studied in \cite{Kasuya:2003va,Enqvist:2003mr,Hamaguchi:2003dc}. In supersymmetric theories there are a number of directions in the field space in which the potential function vanishes called flat directions \cite{Enqvist:2003gh}. The Affleck-Dine (AD) mechanism \cite{Affleck1985361} uses these flat directions for the baryogenesis. It was noted in \cite{Kusenko:1997si,Kasuya:2000wx} that the AD mechanism can lead to  the production of a large number of clumpy scalar field objects called Q-balls. These are a class of non-topological solitons that are frequently generated in different supersymmetric theories of the early universe and have been since studied numerically in two and three dimensional simulations \cite{Kasuya:2000sc,Kasuya:2000wx,Multamaki:2002hv,Hiramatsu:2010dx}.

We will now assume that the Affleck-Dine (AD) field plays the role of the curvaton.
Although this model might not be consistent with current cosmological observations \cite{Hamaguchi:2003dc} we will still use it to study the functionality of the algorithm. The potential function in the gravity-mediated SUSY breaking model is
\begin{equation}
V(\Phi) = m_{\sigma}|\Phi|^2\Big(1 + K\log \Big(\frac{|\Phi|^2}{M_{*}^2} \Big)\Big)- g H^2|\Phi|^2 + \frac{\lambda}{m_{Pl}^2}|\Phi|^6
\end{equation}
where $\Phi$ is the complex curvaton field and the curvaton mass $m_{\sigma}$, $M_{*}$ is the renormalization point, $g$ a constant of order one and $m_{Pl}$ the reduced Planck mass. We will use very similar initial and parameter values that were employed in \cite{Multamaki:2002hv}, namely $m_{\sigma} = 10^2 $ GeV, $M_{*} = 10^{14}$ GeV, $K = -0.1$, $\lambda = 0.5$. Note that the $gH^2$-term can be omitted since $g H^2 \ll m^2_{\sigma}$. The initial values for the fields are set to be
\begin{equation}
\Phi(t_{\textrm{phys},0}) = 2.57 \cdot 10^7 m_{\sigma}, \quad \Phi'(t_{\textrm{phys},0}) =  2.57  \cdot 10^7 m_{\sigma}^2 \; i
\end{equation}
that were used in \cite{Kasuya:2000wx}. We will assume that the universe is dominated by radiation that originates from the inflaton decay. Initial time is set to $t_{\textrm{phys}} = 100/m_{\sigma}$ and the energy density of the homogeneous radiation component is initially $\rho_{\gamma} = 4/(3 t_{\textrm{phys}}^2)$ which sets the initial Hubble parameter value equal to the one used in \cite{Multamaki:2002hv}.
The initial inhomogeneous perturbations are created with the method given in Defrost. We use a $128^3$ lattice with side length set to $L = 8/m_{\sigma}$ and time step $d \eta = 0.005/m_{\sigma}$.

\begin{figure}[t]
\subfigure[]{\label{fig3a}\includegraphics[width=0.45\columnwidth]{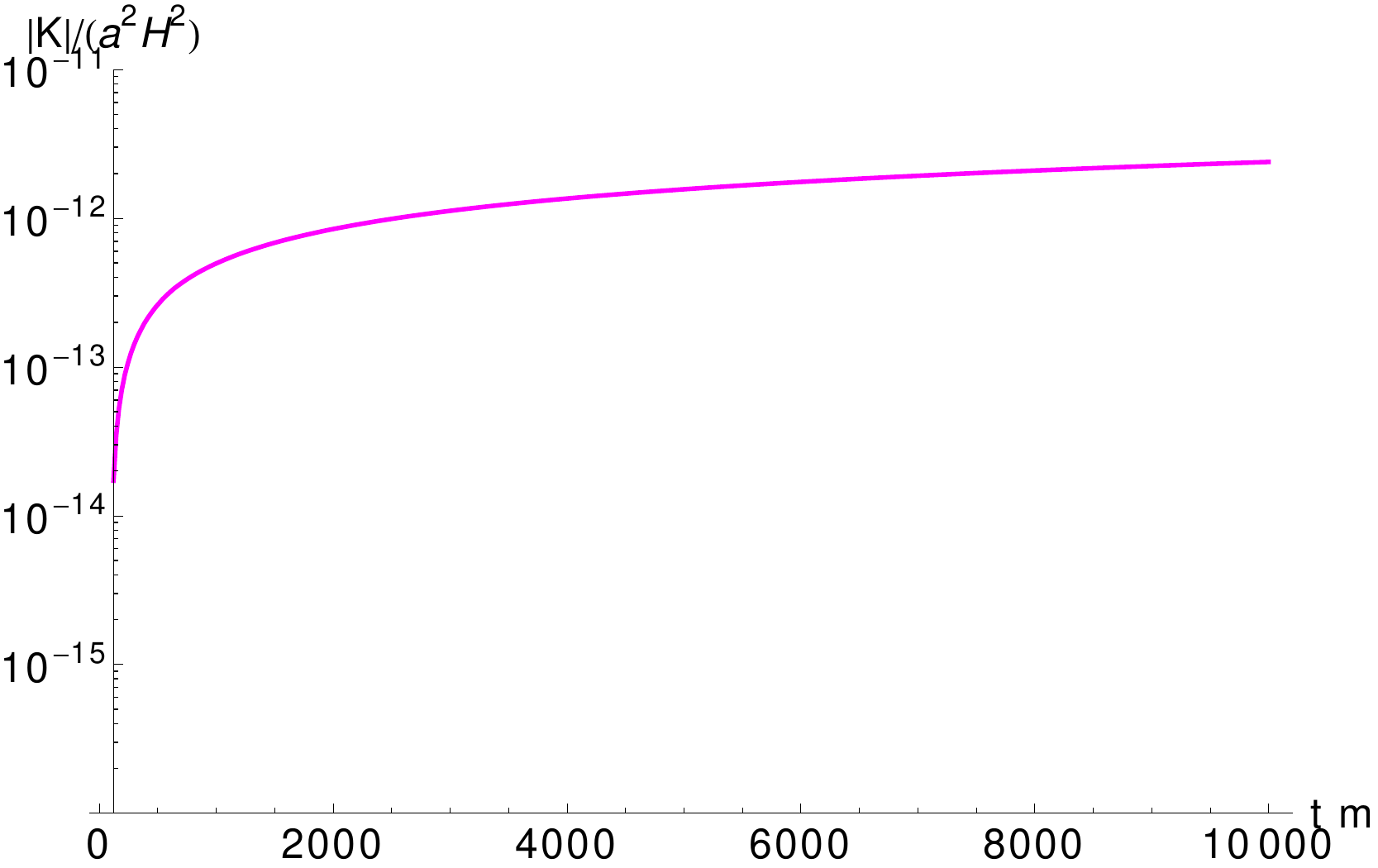}}
\quad
\subfigure[]{\label{fig3b}\includegraphics*[width=0.45\columnwidth]{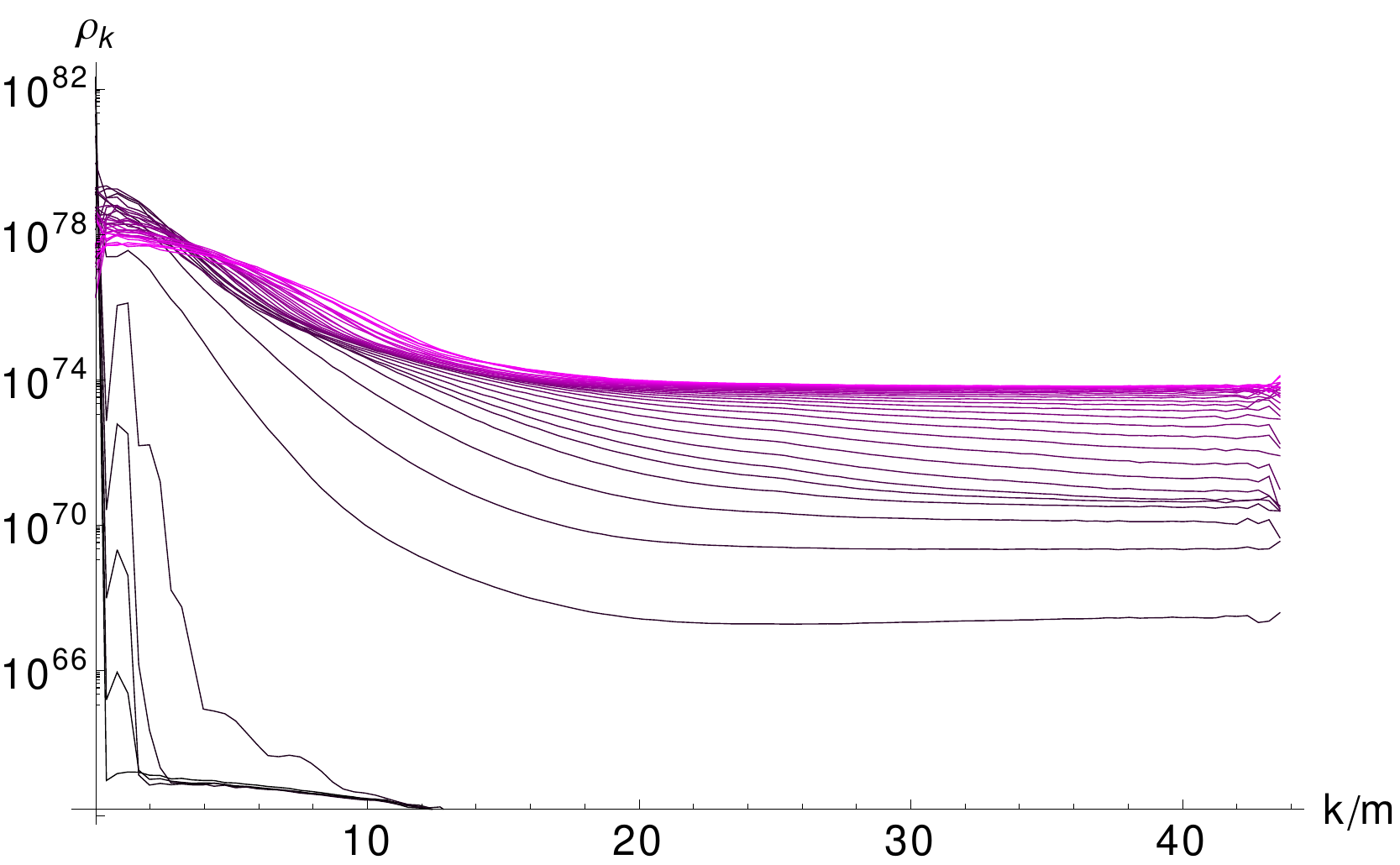}}
\subfigure[]{\label{fig3c}\includegraphics*[width=0.45\columnwidth]{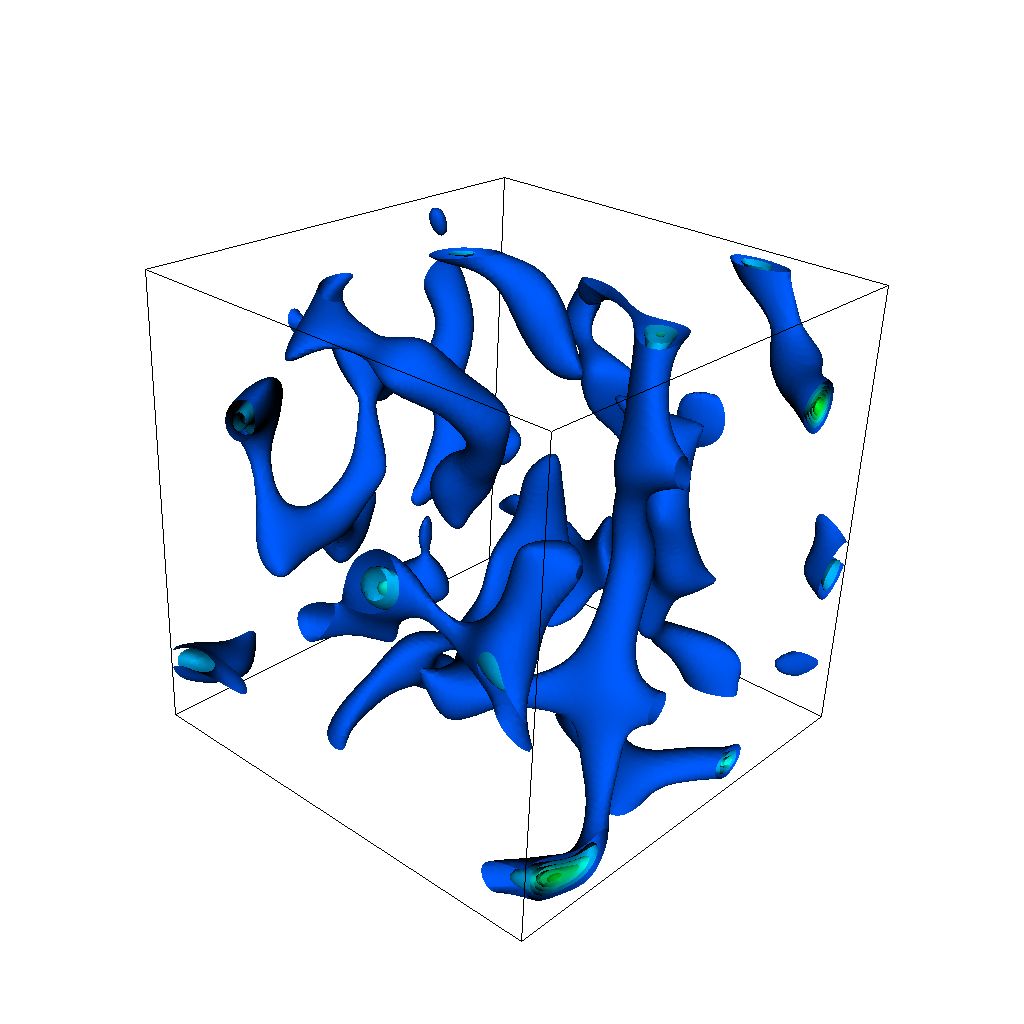}}
\quad
\subfigure[]{\label{fig3d}\includegraphics[width=0.45\columnwidth]{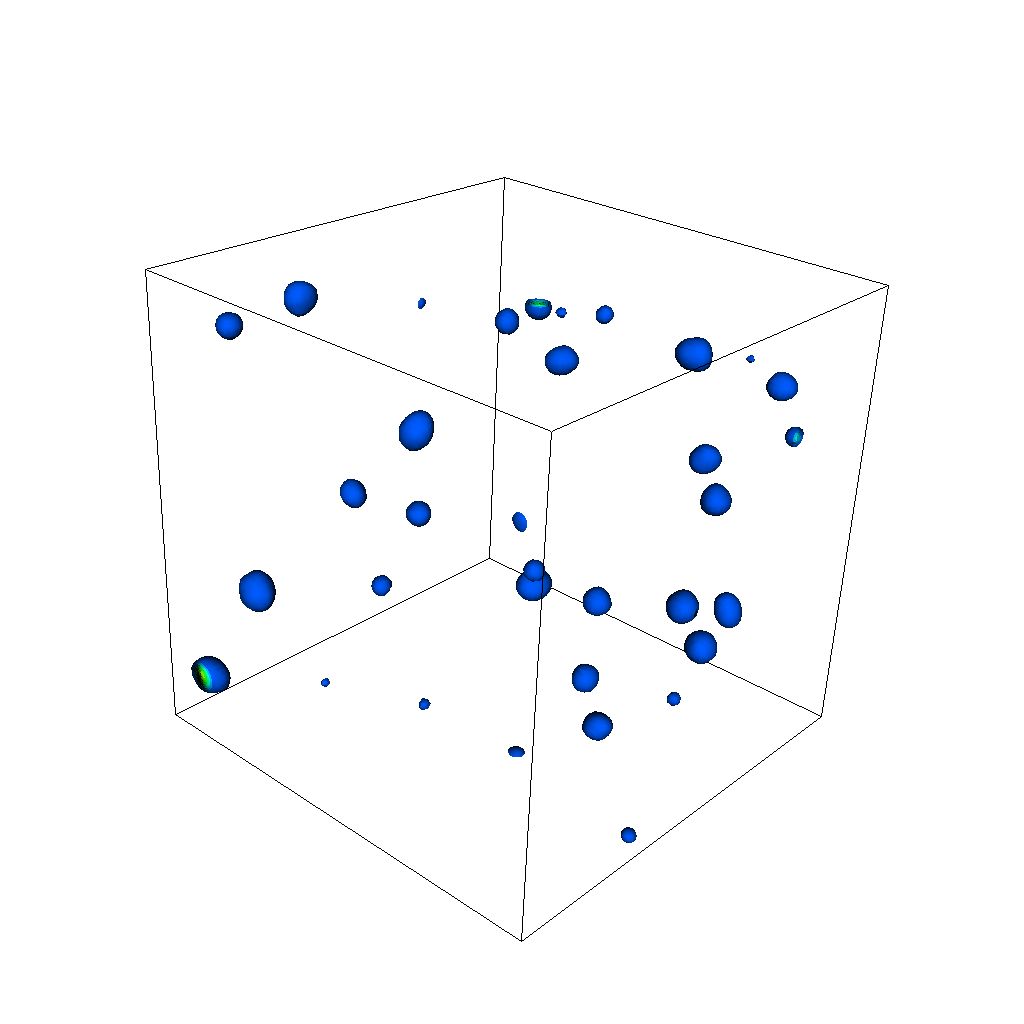}}

\caption{(a) Plot of the relative residual curvature (\ref{eq:res-curv}) as a function of physical time in units of $1/m$. (b) The energy density spectrum of the real component of $\Phi$. Note that the color evolves with time and the lightest curve (at the top) is calculated at $t_\textrm{phys}=10000/m$ whereas darkest one (at the bottom) are evaluated at $t_\textrm{phys}=0/m$. The generation of Q-balls can be seen as a spike in the small $k$ values of the early spectra.
Energy density isosurfaces calculated at (c) $t_{\textrm{phys}}=645.345/m_{\sigma}$ and at (d) $t_{\textrm{phys}}=5088.03/m_{\sigma}$ in the curvaton scenario. The curvaton field clearly fragments into Q-balls.
}
\end{figure}

%

We have plotted the absolute value of the relative residual curvature in figure \ref{fig3a}. The algorithm is seen to be accurate even with a fourth order integrator.
We have also included a plot of energy density spectrum of the real part of the field in figure \ref{fig3c}. The generation of Q-balls can be seen as a clear peak in the infrared modes during the early part of the simulation.
In figures \ref{fig3c} and \ref{fig3d} a snapshot of the energy density of the system with isosurface plots is given at $t_\textrm{phys} = 645.345/m_{\sigma}$ and $t_{\textrm{phys}}=5088.03/m_{\sigma}$ respectively. 
Further study of this system would necessitate the use of post-processing algorithm that identifies the Q-balls inside the lattice \cite{Hiramatsu:2010dx}. This procedure could be also implemented in CUDA language and included with the program but we have however decided to leave this as a possible future upgrade of the program.

These simulations were done with the Tesla card. One simulation takes roughly 2 hours and 33 minutes meaning an average time of $0.032$ seconds per one step. Note that due to the more computationally demanding potential function the code is roughly three times slower than in the oscillon case. A quick test shows also that the consumer card is roughly 60 \% slower than the Tesla version which is mainly due to the difference in computational speed. Simulation of a larger $256^3$ lattice would take roughly 20.5 hours with the Tesla card.

\section{Discussion and conclusions}

We have presented a PyCUDA implementation of symplectic integrator of scalar fields in an expanding space. Although the program has its roots in LATTICEEASY, Defrost and CUDAEASY programs it is based on an entirely new algorithm developed by A. Frolov and Z. Huang \cite{Frolov}. Due to its sympleticity the algorithm is able to conserve the first Friedmann equation to very high precision. We have also further improved and optimized the used method to better suite the GPU. For simulations in which a linearized version of the equations suffice we have derived a linearized Hamiltonian equations and implemented the corresponding symplectic integrator also on the GPU.

Most of the program has been written in Python language to make it as friendly as possible to the end user.
Although Python generally leads to some overhead when compared to a pure C-language implementation and hence a longer runtime the easy of use of the program more than makes up this deficiency. When comparing PyCOOL to previous scalar field integrators the cost of changing to a different scalar field model has been made much lower.
This is based on the textual templating functionality used in PyCOOL (and made possible by PyCUDA and jinja2 library) that creates automatically the simulation code from template files meaning that the end user should only rarely need to edit the actual simulation code. In order to study a new model a simple model file that defines the necessary fields, potential functions and the initial values needs to be created and in most cases the program will handle the rest.

We tested the functionality of PyCOOL with three different models that lead to non-linear dynamics. In all of these we were able to produce results consistent with previous studies. We also tested the accuracy of the program by studying the conservation of the first Friedmann equation. We found that the program was not only able to keep the errors many orders of magnitude smaller compared to older lattice evolution codes but that it can also be orders of magnitude faster than a multithreaded CPU powered code.

To make the program more useful to the end user we have included various postprocessing functions that can be used to improve the understanding of the non-linear process under study. We have illustrated this functionality with selected figures in the numerical results section. Further examples of the output generated by the program will be added to the home page of the program 
\href{http://www.physics.utu.fi/tiedostot/theory/particlecosmology/pycool/}{http://www.physics.utu.fi/tiedostot/theory/particlecosmology/pycool/}.

Although PyCOOL has been done with the intention to make it as good as possible there are several venues were it could be improved. Currently the program uses conformal time in the integration steps. It would be however rather easy to implement also an integrator that uses physical time in the integration steps. This could be also expanded to make it possible to use a more general time variable similar to LATTICEEASY. Other possibilities include a possible pseudo-spectral version to calculate the Laplacian operators. Also with the advent of CUDA 4.0 software multi-GPU implementations have become much easier to manage. A possible multi-GPU version of PyCOOL could either 
solve very large lattices faster or alternatively it could be used to evolve many simulations simultaneously which could be used in non-gaussianity related calculations \cite{Enqvist:2004ey,Chambers:2007se,Chambers:2008gu,Chambers:2009ki} where a large number of lattices with different initial conditions needs to be simulated. A version that uses distributed computing to calculate non-gaussianity would also be an interesting possibility especially when remembering that the performance of the consumer graphics cards was very close to the Tesla cards in the chaotic potential model.

In summary, we believe that PyCOOL has great potential to be very useful to the cosmology and particle physics community when studying different non-linear post-inflationary processes with high precision and speed.

\subsection*{Acknowledgments}

The author would like to thank Andrei Frolov for enlightening help in the symplectic integration procedure. The author is also grateful to Arttu Rajantie for useful comments and for providing a code that was helpful in the linearization process.
Useful comments by Iiro Vilja are also acknowledged.

This paper has been funded by OP Bank Group Research Foundation and the author would like to personally thank Prof. Luis Alvarez for allowing the author to pursue this research at Turku School of Economics.



\end{document}